\documentclass[12pt]{article}
\newlength{\pubnumber} 

\catcode`\@=11
\@addtoreset{equation}{section}

\def\section{\@startsection{section}{1}{\z@}{3.5ex plus 1ex minus .2ex}
 {2.3ex plus .2ex}{\large\bf}}
\def\subsection{\@startsection{subsection}{2}{\z@}{2.3ex plus .2ex}
 {2.3ex plus .2ex}{\bf}}
\newcommand\Appendix[1]{\def\thesection{Appendix \Alph{section}}
 \section{\label{#1}}\def\thesection{\Alph{section}}}
\begin{document}

\begin{titlepage}
\samepage{
\setcounter{page}{0}
\rightline{\tt hep-th/0505233}
\rightline{May 2005}
\vfill
\begin{center}
    {\Large \bf Re-Identifying the Hagedorn Transition\\}
\vfill
   {\large
      Keith R. Dienes\footnote{
     E-mail address:  dienes@physics.arizona.edu}
        $\,$ and $\,$ Michael Lennek\footnote{
     E-mail address:  mlennek@physics.arizona.edu}
    \\}
\vspace{.12in}
 {\it  Department of Physics, University of Arizona, Tucson, AZ  85721  USA\\}
\end{center}
\vfill
\begin{abstract}
  {\rm 
   The Hagedorn transition in string theory is normally
    associated with an exponentially rising density of states,
    or equivalently with the existence of a thermal string winding mode
    which becomes tachyonic above a specific temperature.
   However, the details of the Hagedorn transition
    turn out to depend critically on the precise manner in which  
    a zero-temperature string theory is extrapolated to finite
    temperature.  In this paper, we argue that for broad classes of closed
     string theories, the traditional
     Hagedorn transition is completely absent when the correct
     extrapolation is used.  However, we also argue that there is an alternative
      ``re-identified'' Hagedorn transition which is triggered by the  
    thermal winding excitations of a different, ``effective''
     tachyonic string ground state.
      These arguments allow us to re-identify the Hagedorn temperature
     for heterotic strings.  Moreover, we find that all 
     tachyon-free closed string models in ten dimensions share
     the same (revised) Hagedorn temperature, resulting in a universal Hagedorn
     temperature for both Type~II and heterotic strings.
    We also comment on the possibility of thermal spin-statistics
    violations at the Planck scale.
   }
\end{abstract}
\vfill
\smallskip}
\end{titlepage}

\setcounter{footnote}{0}

\def\beq{\begin{equation}}
\def\eeq{\end{equation}}
\def\beqn{\begin{eqnarray}}
\def\eeqn{\end{eqnarray}}
\def\half{{\textstyle{1\over 2}}}
\def\quarter{{\textstyle{1\over 4}}}

\def\calO{{\cal O}}
\def\calE{{\cal E}}
\def\calT{{\cal T}}
\def\calM{{\cal M}}
\def\calF{{\cal F}}
\def\calS{{\cal S}}
\def\calY{{\cal Y}}
\def\calV{{\cal V}}
\def\ibar{{\overline{\imath}}}
\def\chibar{{\overline{\chi}}}
\def\ttwo{{\vartheta_2}}
\def\tthree{{\vartheta_3}}
\def\tfour{{\vartheta_4}}
\def\ttwob{{\overline{\vartheta}_2}}
\def\tthreeb{{\overline{\vartheta}_3}}
\def\tfourb{{\overline{\vartheta}_4}}

\def\qbar{{\overline{q}}}
\def\mm{{\tilde m}}
\def\nn{{\tilde n}}
\def\rep#1{{\bf {#1}}}
\def\ie{{\it i.e.}\/}
\def\eg{{\it e.g.}\/}

\newcommand{\newc}{\newcommand}
\newc{\gsim}{\lower.7ex\hbox{$\;\stackrel{\textstyle>}{\sim}\;$}}
\newc{\lsim}{\lower.7ex\hbox{$\;\stackrel{\textstyle<}{\sim}\;$}}

\hyphenation{su-per-sym-met-ric non-su-per-sym-met-ric}
\hyphenation{space-time-super-sym-met-ric}
\hyphenation{mod-u-lar mod-u-lar--in-var-i-ant}


\def\inbar{\,\vrule height1.5ex width.4pt depth0pt}

\def\IC{\relax\hbox{$\inbar\kern-.3em{\rm C}$}}
\def\IQ{\relax\hbox{$\inbar\kern-.3em{\rm Q}$}}
\def\IR{\relax{\rm I\kern-.18em R}}
 \font\cmss=cmss10 \font\cmsss=cmss10 at 7pt
\def\IZ{\relax\ifmmode\mathchoice
 {\hbox{\cmss Z\kern-.4em Z}}{\hbox{\cmss Z\kern-.4em Z}}
 {\lower.9pt\hbox{\cmsss Z\kern-.4em Z}}
 {\lower1.2pt\hbox{\cmsss Z\kern-.4em Z}}\else{\cmss Z\kern-.4em Z}\fi}

\long\def\@caption#1[#2]#3{\par\addcontentsline{\csname
  ext@#1\endcsname}{#1}{\protect\numberline{\csname
  the#1\endcsname}{\ignorespaces #2}}\begingroup
    \small
    \@parboxrestore
    \@makecaption{\csname fnum@#1\endcsname}{\ignorespaces #3}\par
  \endgroup}
\catcode`@=12

\input epsf

\section{Motivation, overview, and summary of results}
\setcounter{footnote}{0}

The Hagedorn transition is one of the central hallmarks of string thermodynamics.
Originally discovered in the 1960's through studies of hadronic 
resonances and the so-called ``statistical
bootstrap''~\cite{Hagedorn,Huang,cudell},
the Hagedorn transition is now understood to be a generic feature
of any theory exhibiting a density of states which rises exponentially as
a function of mass.  
In string theory,
the number of states of a given total mass depends on the number of ways
in which that mass can be partitioned amongst 
individual quantized mode contributions, leading to an exponentially rising
density of states~\cite{Polbook}.
Thus, string theories should exhibit a 
Hagedorn transition~\cite{earlystringpapers,vortices,McClainRoth,AtickWitten,longstrings}.
Originally, it was assumed that the Hagedorn temperature is a limiting temperature
at which the internal energy of the system diverges.  However, 
later studies demonstrated that the internal energy actually remains finite at 
this temperature.  This then suggests that the Hagedorn temperature is merely the
critical temperature corresponding to a first- or second-order phase transition.

There have been many speculations concerning possible interpretations of this phase
transition, including a breakdown of the string worldsheet into vortices~\cite{vortices}
or a transition to a single long-string phase~\cite{longstrings}. 
It has also been speculated that there is a 
dramatic loss of degrees of freedom at high temperatures~\cite{AtickWitten}.
Over the past two decades, studies of the Hagedorn transition have reached
across the entire spectrum of modern string-theory research, 
including open strings and D-branes,
strings with non-trivial spacetime geometries (including AdS backgrounds and $pp$-waves), 
strings in magnetic fields, ${\cal N}{=}4$ strings, tensionless
strings, non-critical strings, two-dimensional strings, little strings, matrix models, non-commutative theories, 
as well as possible cosmological implications and implications for the brane world.
A brief selection of papers in many of these areas appears in 
Refs.~\cite{KounnasRostand,general,Dbranes,geometries,ppwaves,magnetic,tensionless,noncritical,little,matrix,NCOS,cosmology,ridge,2Dhet}.
However, with only rare exceptions, the fundamental origins of the Hagedorn transition
have not been seriously investigated within the context of actual finite-temperature
string model-building.

In this paper, we shall undertake a critical re-evaluation of the Hagedorn phenomenon
within the context of perturbative closed string theories.
As we shall show, the details of the Hagedorn
transition --- including its very existence ---
depend on the precise manner in which such zero-temperature string
theories are extrapolated to finite temperature.
Using very general criteria, we shall argue that when a consistent extrapolation
is selected, the traditional Hagedorn transition is completely absent within
a broad class of closed strings consisting of all heterotic strings and certain
Type~II strings in $D<10$.
Indeed, as we shall demonstrate,
the usual Hagedorn phase transition is actually not reflected in the behavior of
 {\it any}\/ string thermodynamic quantities;  
one-loop thermodynamic quantities such as the free energy, the internal energy, the entropy, and even
the specific heat will remain smooth and undisturbed as a function of temperature,
crossing the traditional Hagedorn temperature without so much as a ripple.

So what happened to the Hagedorn transition in such theories?
As we shall argue,
the answer is easy to understand in the heterotic case.
As we shall review in Sect.~3,
one of the standard explanations for the emergence of a Hagedorn
transition~\cite{AtickWitten} involves the appearance of
a thermal winding-mode excitation of the string ground state
which becomes tachyonic beyond a critical (Hagedorn) temperature.
We find, by contrast, that 
there is no Hagedorn divergence in the effective potential of the heterotic
string because
the tachyonic string ground state on which this argument
is predicated is unphysical and does not appear in the
actual string spectrum when a proper finite-temperature string model 
is constructed.
Thus, it is absent from the string partition function,
and does not have any thermal excitations which could give rise
to a divergence.

Let us be more precise here.
There are actually two ways in which a state may
fail to appear in the spectrum of a given closed finite-temperature
string theory:
\begin{itemize}
\item First, a state may fail to satisfy the 
    appropriate finite-temperature GSO constraints.
    Even if it happens to satisfy 
    the level-matching constraints, such a state is unphysical.
    Such states do not appear in the string partition function,
    and play no role in string loop calculations.
\item  Alternatively, a state may satisfy the finite-temperature 
    GSO constraints, but
    fail to satisfy the appropriate level-matching constraints.
    In other words,
    while modular invariance ensures that the difference
    between the left- and right-moving worldsheet energies
    will be an integer, this integer may not be zero.
    Such a state is merely off-shell, since its contributions
    appear in the string partition function.  Such states
    cannot appear as in- or out-states, but they do contribute
    as internal states in string loop diagrams because they
    satisfy the GSO constraints appropriate for the particular
    string model under study.
\end{itemize}

The important point, then, is that the tachyonic heterotic string ground state
on which the usual winding-mode argument is based
is often in the first category --- unphysical rather than merely off-shell.
This means that this state does not exist, except as the
mathematical ground state of the conformal field theory from 
which the physical string states
(both on-shell and off-shell) are constructed.
We shall discuss this more fully in Sect.~4.
Although this state still controls the exponential divergence in the density
of physical string states, it is not itself a physical object.
Thus, even when the mass of this state is augmented by thermal
winding contributions to become massless and level-matched,
this state still fails to satisfy the finite-temperature GSO constraints.
It therefore does not become physical, and cannot trigger a Hagedorn transition.

Is, then, the Hagedorn transition a spurious, unphysical effect in such theories?

In this paper, we shall argue that the answer is ``no'', but that the 
Hagedorn transition has been misidentified.
Instead, we claim that there actually {\it is}\/ a physical
Hagedorn transition, but at a somewhat higher temperature.
Moreover, we claim that this new Hagedorn transition is completely observable
in the behavior of physical, thermodynamic quantities,
appearing as actual divergences or discontinuities
in these quantities
as functions of temperature.
Indeed, we shall show that this new phase transition, 
unlike the traditional Hagedorn transition, 
is an extremely
weak transition whose order depends on the spacetime dimension.

The origins of this new Hagedorn transition can be explained
in a manner identical to the explanation of the traditional Hagedorn
transition.  Specifically, 
we shall show in Sect.~5 that for heterotic
strings there is an
off-shell (but otherwise physical) 
tachyon which is {\it not}\/ the string ground state, but which
    generically appears in all finite-temperature string models.
    When augmented with thermal winding contributions, this
    state can become massless and on-shell.
    This then triggers a true, physical divergence (or discontinuity)
    in the behavior of thermodynamic quantities.
    We claim that for the purposes of calculating
    thermodynamic quantities,
    it is {\it this}\/ state which functions as the ``effective'' ground state:
    it is this state, and not the
    actual string ground state, that is responsible for
    triggering the Hagedorn transition.  Since this effective
    string ground state is less tachyonic than the actual
    string ground state, the corresponding Hagedorn temperature
    is higher than the traditional one.

The above comments apply to heterotic strings.
However, as we shall see, similar remarks will also apply for certain Type~II strings
in dimensions $D<10$.
 
In Sect.~6, we shall then broaden our discussion to investigate the appearance
of other Hagedorn-like transitions.  We shall find that many such additional transitions
may exist, but that they are generally quite model-dependent.

Much of the discussion in this paper will be as general and model-independent as possible.
Consequently, our focus will primarily be on the Hagedorn transition, and not on the
model-dependent issue of determining the correct finite-temperature extrapolation  
of zero-temperature string models.
To compensate for this, in Appendix~A we will explicitly analyze the case of the ten-dimensional
supersymmetric $SO(32)$ heterotic string.
We shall explicitly construct what we believe is the finite-temperature extrapolation
for this theory, and in the process lay out our general criteria for such extrapolations.
Moreover, using these criteria, we shall show that each of the tachyon-free ten-dimensional
heterotic strings experiences a re-identified Hagedorn
transition at a temperature normally associated with Type~II strings.  
This includes not only the supersymmetric $SO(32)$ and $E_8\times E_8$ 
heterotic strings, but also the ten-dimensional tachyon-free non-supersymmetric 
$SO(16)\times SO(16)$ heterotic string~\cite{AGMV,DH}.
Thus, we conclude that {\it all tachyon-free ten-dimensional closed strings
actually have a universal Hagedorn temperature}. 
We shall also comment on the possibility of thermal spin-statistics
violations at the Planck scale, and show that this possibility is intimately
connected with the re-identification of the Hagedorn phenomenon.

\section{Calculating thermodynamic potentials in string theory}
\setcounter{footnote}{0}
 
We begin by quickly reviewing 
the calculation of one-loop thermodynamic quantities in closed string theories.
Our purpose here is merely to 
recall established formalism
and set notational conventions.  For detailed derivations
or explanations, we refer the reader to the original 
literature or Ref.~\cite{Polbook}.
 
Just as in ordinary statistical mechanics, the fundamental
quantity of interest in string thermodynamics is
the one-loop thermal string partition function $Z_{\rm string}(\tau,T)$.
This is 
a function of not only the temperature $T$ but also 
the complex modular parameter $\tau$ (parametrizing the shape of the one-loop
toroidal string worldsheet).  In terms of $Z_{\rm string}(\tau,T)$, the one-loop 
thermal vacuum amplitude is then given by the one-loop modular integral
\beq
    \calV(T) ~\equiv~ -\half \,{\cal M}^{D-1}\, \int_{\cal F} 
              {d^2 \tau\over ({\rm Im} \,\tau)^2}
             \,Z_{\rm string}(\tau,T)
\label{Vdef}
\eeq
where ${\cal M}\equiv M_{\rm string}/(2\pi)$ is the reduced string scale;
$D$ is the number of non-compact spacetime dimensions;
and where
\beq
   {\cal F}~\equiv ~\lbrace \tau:  ~|{\rm Re}\,\tau|\leq \half,
 {\rm Im}\,\tau>0, |\tau|\geq 1\rbrace
\label{Fdef}
\eeq
is the fundamental domain
of the modular group.
For future notational convenience, we shall define $\tau_1\equiv {\rm Re}\,\tau$ and
$\tau_2\equiv {\rm Im}\,\tau$.
In general, $\calV(T)$ plays the role usually taken by the logarithm of the
statistical-mechanical partition function.  
Given this definition for $\calV$, the free energy $F$, internal
energy $U$, entropy $S$, and specific heat $c_V$ then follow
from the standard thermodynamic definitions:
\beq
          F = T \calV~,~~~~
         U = - T^2 {d\over dT} \calV~,~~~~
         S = -{d\over dT} F~,~~~~ c_V = {d\over dT} U~.
\label{usualrelations}
\eeq

Because of its central role in determining the thermodynamics
of the corresponding string theory, we shall now focus on the
calculation of the string thermal partition function $Z_{\rm string}(\tau,T)$.
Let us begin by discussing the case of a compactified bosonic string  
at zero temperature.
In such a case, we have
\beq
      Z_{\rm model}(\tau)~\equiv~ {\rm Tr} ~ \overline{q}^{H_R}\, q^{H_L}
\label{PF}
\eeq
where the trace is over the complete Fock
space of states in the theory.
Here $q\equiv \exp(2\pi i \tau)$, and $(H_R,H_L)$
denote the worldsheet energies for the right- and left-moving worldsheet
degrees of freedom, respectively.
For example,
in the case of the bosonic string compactified
to $D$ spacetime dimensions, $Z_{\rm model}$ takes the
general form
\beq
         Z_{\rm model}(\tau) ~=~ \tau_2^{1-D/2}\,
          \, {\overline{\Theta}^{26-D} \Theta^{26-D}
         \over \overline{\eta}^{24} \eta^{24} }
\label{Zstringgenform}
\eeq
where the numerator
$ \overline{\Theta}^{26-D} \Theta^{26-D}$
schematically represents a sum over the $2(26-D)$-dimensional
compactification lattice for left- and right-movers.
Note that in general, $Z_{\rm model}$ is the quantity which appears
in the calculation of the one-loop cosmological
constant (vacuum energy density) of the model:
\beq
         \Lambda ~\equiv~
     -\half \,{\cal M}^D\, \int_{\cal F} {d^2 \tau\over \tau_2^2}
             Z_{\rm model}~
\label{Lambdadef}
\eeq
Of course, 
this quantity is divergent 
for the compactified bosonic string
as a result of the physical bosonic-string tachyon.

In order to extend such a model to finite temperature, we
recall that
in string theory (just as in ordinary quantum field theory),
finite-temperature effects can be
incorporated~\cite{Polbook,Pol86}
by compactifying an extra (Euclidean)
time dimension on a circle
of radius $R_T = (2\pi T)^{-1}$.
The Matsubara modes are nothing but the Kaluza-Klein states corresponding
to this compactification.
Since all of the states in these string
models are presumed to be bosonic, and since each bosonic state must be assigned
periodic boundary conditions around this extra dimension, we see that
the Matsubara frequencies for each of the zero-temperature string states
have integer modings.
However, for extended objects such as closed strings,
we must include not only
``momentum'' Matsubara states
(as described above),
but also ``winding'' Matsubara states
(analogues of the usual string winding modes).
Both types of states are necessary for the modular invariance of the
underlying theory at finite temperature.
We thus obtain a full, thermal partition function
of the form
\beq
         Z_{\rm string}(\tau,T) ~\equiv~
         Z_{\rm model}(\tau) \, Z_{\rm circ}(\tau,T)~
\label{parfunct}
\eeq
where the extra factor $Z_{\rm circ}$
represents a double summation over integer Matsubara momentum and winding modes:
\beq
     Z_{\rm circ}(\tau,T)~=~
     \sqrt{ \tau_2}\,
    \sum_{m,n\in\IZ} \,
      \overline{q}^{(ma-n/a)^2/4}  \,q^{(ma+n/a)^2/4}
\label{Zcircdef}
\eeq
with $a\equiv 2\pi T/M_{\rm string} \equiv T/\calM$.
It is the full thermal partition function $Z_{\rm string}(\tau,T)$ which
is then used in the calculation of the vacuum amplitude in Eq.~(\ref{Vdef}).
Note that $Z_{\rm circ}\to 1/a$ as $a\equiv T/\calM \to 0$.  We therefore
find that $\calV(T)\to \Lambda/T$, and thus $F(T)\to \Lambda$, as $T/\calM\to 0$.
By contrast, we have $\calV(T)\to \Lambda T/\calM^2$ 
as $T/\calM\to \infty$, 
implying that $F(T)\to \Lambda T^2/\calM^2$.

Let us now proceed to discuss the more general case of
fermionic string theories (such as the superstring
and the heterotic string).  The critical differences relative
to the bosonic string are the presence of spacetime fermions
in the spacetime spectrum and the possibility of removing
on-shell tachyons through non-trivial GSO projections.

Once again, let us begin by considering the zero-temperature theory.
The partition function for such a theory takes the form
\beq
      Z_{\rm model}(\tau)~\equiv~ {\rm Tr}\,
          (-1)^F\, \overline{q}^{H_R}\, q^{H_L}~,
\label{PF2}
\eeq
which is completely analogous to Eq.~(\ref{PF}) except for the
spacetime statistics factor $(-1)^F$, where $F$ is the spacetime fermion number.
Thus bosonic states contribute positively to $Z_{\rm model}$, while
fermionic states contribute negatively.
Given this,
the one-loop cosmological constant is given by the same one-loop integral
in Eq.~(\ref{Lambdadef});  however, the presence of non-trivial
GSO projections can now eliminate on-shell tachyons and result in a finite 
cosmological constant.

In order to extend this theory to finite temperature, we again
must introduce the summations over thermal Matsubara states.
However, it is here that the primary difference arises:
while bosonic states must be periodic around the extra
(Euclidean) time direction, resulting in integer-moded Matsubara
frequencies, the fermionic states must be {\it antiperiodic}\/
around this direction, resulting
in Matsubara modings which are {\it integer plus one-half}\/.
Thus, the bosonic and fermionic portions of $Z_{\rm model}$ must
be multiplied by different Matsubara sums, destroying the simple
factorized form in Eq.~(\ref{parfunct}).
In general, this structure can also be further complicated by subsequent
orbifolding and the introduction
of temperature-dependent Wilson lines.
In such cases, modular invariance can serve as a useful tool for constraining
the form of the resulting partition 
functions~\cite{Rohm,AlvOso,AtickWitten,KounnasRostand}, 
but other physical
criteria (such as proper thermal spin-statistics) play an important role.

Towards this end, let us
introduce~\cite{Rohm} four new functions $\calE_{0,1/2}$ and $\calO_{0,1/2}$
which are the same as the summation in
$Z_{\rm circ}$ in Eq.~(\ref{Zcircdef}) except for the following
restrictions on their summation variables:\footnote{Note that these functions
are to be distinguished from a related (and also often used) 
set of functions with the same names in which the roles of $m$ and $n$ are exchanged.}
\beqn
       \calE_0 &=& \lbrace  m\in\IZ,~n~{\rm even}\rbrace\nonumber\\
       \calE_{1/2} &=& \lbrace  m\in\IZ+\half ,~n~{\rm even}\rbrace\nonumber\\
       \calO_0 &=& \lbrace  m\in\IZ,~n~{\rm odd}\rbrace\nonumber\\
       \calO_{1/2} &=& \lbrace  m\in\IZ+\half ,~n~{\rm odd}\rbrace~.
\label{EOfunctions}
\eeqn
Under the modular transformation $T:\tau\to\tau+1$,
the first three functions are invariant while
$\calO_{1/2}$ picks up a minus sign;
likewise, under $S:\tau\to-1/\tau$,
these functions mix according to
\beq
   \pmatrix{  \calE_0 \cr \calE_{1/2} \cr \calO_0  \cr  \calO_{1/2}  } (-1/\tau) ~=~
   \half \pmatrix{   1 & 1 & 1 & 1 \cr
                   1 & 1 & -1 & -1 \cr
                   1 & -1 & 1 & -1 \cr
                   1 & -1 & -1 & 1 \cr}
   \pmatrix{  \calE_0 \cr \calE_{1/2} \cr \calO_0  \cr  \calO_{1/2}  } (\tau) ~.
\label{Smixing}
\eeq
Note that in the $T/\calM\to 0$ limit,
$\calO_0$ and $\calO_{1/2}$ each vanish while $\calE_0,\calE_{1/2} \to \calM/T$;
by contrast, as $T/\calM\to\infty$,
$\calE_{1/2}$ and $\calO_{1/2}$ each vanish while $\calE_0,\calO_0 \to T/(2\calM)$.
Clearly, $\calE_{0}+\calO_{0}= Z_{\rm circ}$.

In terms of these functions,
our complete thermal string partition function for fermionic string theories
then generically takes the form~\cite{Rohm,AlvOso,AtickWitten,KounnasRostand} 
\beqn
        Z_{\rm string}(\tau,T)  &=&
           Z^{(1)}(\tau) ~ \calE_0(\tau,T) ~+~
           Z^{(2)}(\tau) ~ \calE_{1/2}(\tau,T) \nonumber\\
           && ~~+~  Z^{(3)}(\tau) ~ \calO_{0}(\tau,T) ~+~
           Z^{(4)}(\tau) ~ \calO_{1/2}(\tau,T) ~.
\label{EOmix}
\eeqn
Clearly, the individual blocks $Z^{(i)}$ must transform 
under modular transformations in such a way that $Z_{\rm string}$
is modular invariant;  in this case, this implies that each $Z^{(i)}$ must
transform exactly as does its corresponding $\calE/\calO$ function.
Note that the original corresponding zero-temperature partition function is given by
\beq
          Z_{\rm model} ~=~ Z^{(1)} + Z^{(2)}~,
\label{origmodel}
\eeq
since these
are the only two terms which survive the $T/\calM\to 0$ limit.
We thus continue to find the asymptotic behavior $\calV(T)\to \Lambda/T$
as $T/\calM\to 0$. 

It is interesting to note that the opposite limit as $T/\calM\to \infty$
results in the behavior $V(T)\to \tilde \Lambda T/(2\calM^2)$ and 
$F(T)\to \tilde \Lambda T^2/(2\calM^2)$ where $\tilde \Lambda$ is the
cosmological constant associated with the alternate
zero-temperature model whose partition function is given by
\beq
   \tilde Z_{\rm model} ~=~ Z^{(1)} + Z^{(3)}~.
\label{tildeZ}
\eeq
Thus, the thermal partition function (\ref{EOmix}) can be viewed
as mathematically {\it interpolating}\/ between one zero-temperature string model at $T=0$ 
and a {\it different}\/ zero-temperature string model as $T\to\infty$.
The extra factor of two in the asymptotic behavior in the infinite-temperature
limit arises from the reduction of the volume of the $\IZ_2$ orbifold 
that implements the appropriate thermal twists for the spacetime fermions.

Note that the $\calE/\calO$ functions satisfy the
identities
\beqn
     \calE_0(1/a) ~=~ \calE_0(2a)~,~~~~~
    && \calE_{1/2}(1/a) ~=~ \calO_0(2a)~\nonumber\\
    \calO_{0}(1/a)~=~ \calE_{1/2}(2a)~,~~~~~
    && \calO_{1/2}(1/a) ~=~ \calO_{1/2}(2a)~
\label{ident}
\eeqn
where $a=T/\calM$.
Thus, for every partition function of the form in Eq.~(\ref{EOmix}),
there is another in which we replace $a\to 2/a$ and exchange 
$Z^{(2)}$ and
$Z^{(3)}$.
This has the net effect of preserving the interpolation, but exchanging
the $T\to 0$ and $T\to\infty$ limits.

We see, then, that a zero-temperature model whose partition function 
is given in Eq.~(\ref{origmodel}) will have a finite-temperature extrapolation
of the form in Eq.~(\ref{EOmix}).  It is important to note, however, that the
particular form of Eq.~(\ref{EOmix}) is not uniquely determined simply
by the zero-temperature partition function sum $Z_{\rm model}$ 
in Eq.~(\ref{origmodel});  it also depends on how
$Z_{\rm model}$ is divided into $Z^{(1)}$ and $Z^{(2)}$.  
In other words, modular invariance alone is not sufficient to 
determine the unique finite-temperature extrapolation of a given 
zero-temperature model unless one assumes that all bosonic states in $Z_{\rm model}$
are part of $Z^{(1)}$ and all fermionic states are part of $Z^{(2)}$.
As mentioned above, this simple assumption may be affected by 
orbifold twists and string consistency constraints. 
We shall discuss this issue further in Appendix~A.

Finally, it is also easy to see from Eq.~(\ref{EOmix}) why thermal effects
break spacetime supersymmetry.  In the $T\to 0$ limit, our partition function
reduces to $Z_{\rm model}= Z^{(1)} + Z^{(2)}$;  
if this limit is supersymmetric, 
then we necessarily have $Z^{(1)} = -  Z^{(2)}$ 
as an identity on the $q$-expansions of these expressions, resulting in the zero-temperature
supersymmetric limit $Z_{\rm model}=0$, with $\Lambda=0$.
However, even though this partition function vanishes at zero temperature,
this cancellation will not persist at non-zero temperatures because   
$Z^{(1)}$ and 
$Z^{(2)}$ are multiplied by different $\calE/\calO$ functions
representing the traces over 
different thermal Matsubara and/or winding modes. 
This is the reflection of the fact that thermal effects necessarily treat
bosons and fermions differently, thereby breaking any supersymmetry which
may have existed at zero temperature.

\section{The usual Hagedorn transition:  Standard arguments} 
\setcounter{footnote}{0}

Our main concern in this paper is to demonstrate that the traditional
Hagedorn phenomenon is non-existent for a wide class of 
closed strings (including all heterotic strings), and should actually be re-identified.
Let us therefore first review the standard arguments for the appearance
of the Hagedorn transition.  In Sect.~4, we shall then discuss why this transition is,
in fact, absent.

\subsection{The UV argument}

The usual ultraviolet (UV) argument for the existence
of a Hagedorn temperature involves the exponential rise in the number
of string states as a function of mass.
In a nutshell,  if $g_M$
denotes the number
of states with mass $M$, then the thermal partition function is
given by $Z(T)=\sum  g_M e^{-M/T}$.  However, if $g_M \sim e^{\alpha M}$ as $M\to\infty$, then
$Z(T)$ diverges for $T\geq 1/\alpha$.   This
suggests the existence of some sort of phase
transition at the critical (Hagedorn) temperature $T_H\equiv 1/\alpha$.

Let us now develop this argument 
in the precise language of our string partition functions.
We begin by considering the zero-temperature
(\ie, non-thermal) partition function $Z_{\rm model}$.
Since this partition function represents a trace over the string Fock space
as in Eq.~(\ref{PF2}), it encodes the information about the net degeneracies
of string states at each mass level in the zero-temperature theory.  Specifically, if we expand
$Z_{\rm model}$ as a power series in $q$ and $\qbar$, we obtain 
an expression of the form\footnote{We designate our
summation variables as $\mm$ and $\nn$ in order to distinguish them from the thermal
momentum and winding momentum modes $m$ and $n$ in 
Eqs.~(\protect\ref{Zcircdef}) and (\protect\ref{EOfunctions}).}
\beq
        Z_{\rm model}(\tau) ~=~  \tau_2^{1-D/2} \, 
       \sum_{\tilde m,\tilde n}\, a_{\tilde m \tilde n} \,\qbar^{\tilde m} \,q^{\tilde n}~.
\label{bare}
\eeq
Here $(\mm,\nn)$ represent the possible eigenvalues of the right- and left-moving worldsheet
Hamiltonians $(H_R,H_L)$, and $a_{\mm\nn}$ represents the number of bosonic
minus fermionic states which actually have those eigenvalues and satisfy the GSO constraints.  
Modular invariance requires that $\mm-\nn\in \IZ$ for all $a_{\mm\nn}\not=0$;
a state is said to be ``on-shell'' or ``level-matched'' if $\mm=\nn$.
Note that the $\tau_2^{1-D/2}$ prefactor represents the result of the
integration (\ie, the trace) over the continuous spectrum of states corresponding 
to the uncompactified dimensions.  

In general, the spacetime mass of an arbitrary $(\mm,\nn)$ state is
given by
\beq 
          \alpha' M_{\mm\nn}^2 ~=~  2(\mm+\nn)~ 
\label{spacetimemass}
\eeq
where $\alpha'=1/M_{\rm string}^2$. 
Thus, we see that states for which $\mm+\nn\geq 0 $ are massive and/or massless,
while states with $\mm+\nn <0$ are tachyonic.
In general, one constructs a consistent string model in such a way
as to avoid on-shell tachyons (\ie, to ensure that $a_{\mm\nn}=0$ for all $\mm=\nn<0$).  
Of course, if the 
model exhibits spacetime supersymmetry at zero temperature, 
then $a_{\mm\nn}=0$ for all $\mm,\nn$, but the supersymmetry will be broken by 
finite-temperature effects which eventually multiply the bosonic and fermionic
contributions to $Z_{\rm model}$ with different thermal $\calE/\calO$ functions.  
Therefore, for the purposes of the traditional Hagedorn derivation,
we shall consider the bosonic and fermionic
contributions to $a_{\mm\nn}$ separately.

In order to discuss the behavior of $a_{\mm \nn}$, let us first recall that
$Z_{\rm model}$ can generally be constructed in terms of the
characters $\chi_i$ and $\overline{\chi}_\ibar$ of the corresponding left- and right-moving 
conformal field theories (CFTs) on the string worldsheet:
\beq
      Z_{\rm model}(\tau) ~=~ \tau_2^{1-D/2}\, \sum_{\ibar i}\,  
          \overline{\chi}_\ibar (\overline{\tau}) \, N_{\ibar i} \, \chi_i(\tau)~.
\label{Zchi}
\eeq
The coefficients $N_{\ibar i}$ describe the manner in which the left- and right-moving
CFTs are stitched together, and encode the information concerning the GSO
projections inherent in the given string model.
In general, each character $\chi_i$ represents a trace over the sector of the CFT
corresponding to a specific primary field $\phi_i$,
and has a $q$-expansion of the form
\beq
      \chi_i(\tau) ~=~   q^{h_i-c/24} \, \sum_{p=0}^\infty\,  a_p^{(i)} \, q^p ~,
\label{chars}
\eeq
where $c$ is the central charge of the CFT and
$h_i$ is the conformal weight of the primary field $\phi_i$.
Moreover, the coefficients $a_p^{(i)}$
count the number of descendent fields, and are known to grow asymptotically
as~\cite{HR,Kani,missusy}
\beq
       a_p^{(i)} ~\sim~ \exp\left( 4\pi \sqrt{c p\over 24} \right)~ ~~~~~{\rm as}~
               p\to\infty~.
\eeq
Note that this rate of asymptotic growth 
applies for all sectors of the CFT.

Thus, combining our results for the left- and right-movers, we see that 
the numbers of on-shell bosonic or fermionic states in a given string theory generally 
grow as
\beq
      a_{\nn\nn}~\sim~ \exp\left[ 4\pi 
         \left( \sqrt{c_L \over 24} + \sqrt{c_R\over 24}\right) \sqrt{\nn} \right]~ ~~~~~{\rm as}~
               \nn\to\infty~,
\label{asymgrowth}
\eeq
where $c_{L,R}$ are the central charges of the left- and right-moving worldsheet CFTs
in light-cone gauge.
Note that in this result, we have set $\mm=\nn$ because we are focusing on the
on-shell states exclusively.

Eq.~(\ref{asymgrowth}) describes the rate of exponential growth in the numbers of
physical
bosonic and/or fermionic
string states in terms of the central charges $c_{L,R}$ of the left- and right-moving
worldsheet CFTs in light-cone gauge.
As such, this result is completely general, and applies to all bosonic strings,
superstrings, or heterotic strings.

Using this result, the standard argument for the Hagedorn transition proceeds
by taking this collection of states and calculating the thermal partition function 
as we would in ordinary point-particle theories.  Specifically, we substitute the
degeneracies $a_{\nn\nn}$ from Eq.~(\ref{asymgrowth}) into a general partition
function of the form
\beq
          Z ~=~ \sum_\nn  \, a_{\nn\nn} \,e^{-M_{\nn\nn}/T}
\label{wrongZ}
\eeq
where $T$ is the temperature
and  where $M_{\nn\nn}= 2 \sqrt{\nn} M_{\rm string}$ is the spacetime mass 
corresponding to the $\mm=\nn$ level, as given in Eq.~(\ref{spacetimemass}).
We thus immediately find that this partition function will diverge for temperatures
$T\geq T_H$, where
\beq
       T_H ~=~  {1\over 2\pi} \left(  \sqrt{c_L\over 24} + \sqrt{c_R\over 24}
              \right)^{-1}\, M_{\rm string}~.
\label{THag}
\eeq
This is therefore identified as the Hagedorn temperature.
For the bosonic string, we have the light-cone gauge central charges
$(c_L,c_R)=(24,24)$, while
for superstrings (Type II strings) we have
$(c_L,c_R)=(12,12)$ and for 
heterotic strings we have
$(c_L,c_R)=(24,12)$.
We thus find that 
\beq
    {T_H\over M_{\rm string}} ~=~ \cases{
         (4\pi)^{-1}  &  for the bosonic string,\cr
        (2\sqrt{2}\pi)^{-1} &  for the Type II superstring, \cr
        [(2+\sqrt{2})\pi]^{-1} & for the heterotic string.\cr}
\label{Hagtemps}
\eeq
These are indeed the traditional Hagedorn temperatures normally associated
with these theories.

\subsection{The IR argument}

The alternative, infrared (IR) way of understanding the emergence of the Hagedorn transition
involves the {\it low}\/-lying string states. 
As we have seen, the lowest-lying string state is, {\it a priori}\/, 
the tachyonic string ground state.   This in turn arises
from the identity sectors of the respective right- and left-moving worldsheet 
CFTs, with $h_\ibar=h_i=0$.  Thus, according to Eq.~(\ref{chars}),
the string ground state has right- and left-moving
worldsheet energies
\beq
   (H_R,H_L) ~=~ \left(  -{c_R \over 24}~,~ -{c_L\over 24}\right)~.
\label{groundstate}
\eeq

While this state is necessarily tachyonic at zero temperature, 
at non-zero temperatures
it gives rise to an infinite tower of associated thermal momentum and winding
modes.
As a result of 
additional thermal mass contributions that result from such momenta and windings, 
such thermal states can be  massless
or even massive.
The magnitudes of these mass contributions are generally encoded 
within the $\calE/\calO$ functions
which, according to Eq.~(\ref{Zcircdef}), correspond to the 
additional worldsheet energies
\beq
   ({\Delta H}_R,\Delta H_L) ~=~ \left[ \quarter\, (ma -n/a)^2~,
       ~\quarter\, (ma+n/a)^2   \right]~
\label{thermcontribs}
\eeq
where $a\equiv T/\calM$ and 
where $(m,n)$ are the thermal momentum and winding numbers. 
Thus, in order for one of the thermal excitation modes of the string ground state 
to become massless, we must
satisfy the constraints
\beqn
              ma + n/a &=& \pm 2 \sqrt{c_L/24}~ \nonumber\\
              ma - n/a &=& \pm 2 \sqrt{c_R/24}~ 
\label{twoconstr}
\eeqn
where the two $\pm$ signs are uncorrelated. 
Taking the difference of these two constraint equations as well as the
difference of their squares then yields:
\beq
        n/a ~=~ \pm \left( \sqrt{c_L\over 24} \pm \sqrt{c_R\over 24}\right)~,
    ~~~~~~mn ~=~ {c_L-c_R\over 24}~, 
\label{mnconstraint}
\eeq
where again the signs in the first equation are uncorrelated.  

Since the $(m,n)$ quantum numbers are also subject to the restrictions
appropriate for the particular $\calE/\calO$ functions listed in Eq.~(\ref{EOfunctions}),
we see that the second constraint in Eq.~(\ref{mnconstraint}) has only  
two solutions with non-zero
thermal winding number
for each string:
\beqn
      {\rm bosonic},~{\rm Type~II:}~~~~~  && m=0~,~~n=\pm 1~,\nonumber\\
      {\rm heterotic:}~~~~~  && m=\pm \half~,~~n=\pm 1~,
\label{mnsolns}
\eeqn
where the signs in the second line are now correlated.
Thus, inserting $n=\pm 1$ into the first constraint in Eq.~(\ref{mnconstraint}),
we find that there are generally only
two positive solutions for $a$:
\beq
          a^{(\pm)} ~=~ \left| \sqrt{c_L\over 24} \pm \sqrt{c_R\over 24} \right|^{-1}~.
\eeq 
These correspond to the temperatures
\beq
          T^{(\pm)} ~=~ {1\over 2\pi} 
          \left| \sqrt{c_L\over 24} \pm \sqrt{c_R\over 24} \right|^{-1}\, M_{\rm string}~.
\label{Tplusminus}
\eeq
Note that $T^{(+)}< T^{(-)}$.

It is easy to interpret these equations physically.    
At small temperatures, the thermal excitation of the string 
ground state in Eq.~(\ref{mnsolns})
is extremely massive.
This is a direct result of the non-trivial thermal winding number in Eq.~(\ref{mnsolns}), 
which provides a huge mass contribution at small temperatures.
However, as the temperature increases, 
this additional mass contribution becomes smaller and smaller until
eventually this state becomes massless.  This occurs at the temperature $T^{(+)}$,
which is of course in complete agreement with the Hagedorn temperature obtained
in Eq.~(\ref{THag}).  
Thus, we see that the transition triggered by the appearance of the new massless
state at $T^{(+)}$ is nothing but the Hagedorn transition.
Indeed, formally continuing beyond $T^{(+)}$, we would find that 
this state becomes tachyonic, in agreement with our expectations
of a phase transition.  

It is also possible to interpret $T^{(-)}$ in the case of the heterotic string.
If we were to continue to formally increase the temperature beyond $T^{(+)}$,
this tachyonic state would ultimately reach a 
maximum depth before
turning around and becoming more massive again
as a result of the non-zero 
heterotic thermal {\it momentum}\/ mode with $m=\pm 1/2$. 
This state
would ultimately become massless again at the higher temperature $T^{(-)}$.
In the case of the bosonic or Type~II strings, by contrast, 
our Hagedorn state has a vanishing thermal winding number.
It therefore becomes increasingly tachyonic as we
increase the temperature beyond $T^{(+)}$, ultimately settling
into the original tachyonic ground state as $T\to \infty$.
Thus, in such cases, we see that $T^{(-)}$ is essentially infinite.  
Of course, these are only formal interpretations of the above equations
since we should not extrapolate
our calculation beyond the Hagedorn transition at $T^{(+)}$.

Note that in all cases, the Hagedorn temperature $T^{(+)}$ calculated through this tachyonic
winding-mode argument agrees with the Hagedorn temperature calculated through
the exponential density of states argument in Eq.~(\ref{THag}).  
As we can see from the above derivations,
this agreement arises because
the rate of exponential growth
of the density of states given in Eq.~(\ref{asymgrowth})
is correlated with the
depth of the original tachyonic string ground state
given in Eq.~(\ref{groundstate}),
with both quantities set by the central charge of the
string worldsheet theory.

\section{ Why the standard Hagedorn arguments fail} 
\setcounter{footnote}{0}

So what fails in the above arguments?  
In this section, we shall describe why the above Hagedorn transition
is, in fact, completely invisible in calculations of the standard thermodynamic
quantities.  We shall concentrate on the case of the heterotic
string, deferring our discussion of Type~II strings until the end of this
section.

\subsection{The UV argument}

Let us begin by discussing the ultraviolet derivation based on the 
exponentially rising density of states. 
As we have seen in Eq.~(\ref{asymgrowth}), the numbers of
on-shell bosonic and fermionic states 
in a given string theory generically grow
exponentially with the spacetime mass.
In Eq.~(\ref{wrongZ}), these degeneracies are inserted into a 
point-particle partition function, leading to a divergence above the
Hagedorn temperature.  
The problem, of course, is that the partition
function in Eq.~(\ref{wrongZ}) is not a proper string-theoretic
partition function;  it assumes that the string is nothing but a 
collection of the states to which its excitations give rise.
Instead, however, we must do a proper string-theoretic vacuum-amplitude calculation
as outlined in Eq.~(\ref{Vdef}), using a string partition function which
depends not only on the temperature $T$ but also a torus parameter $\tau$.

There are three major differences between the point-particle partition
function in Eq.~(\ref{wrongZ}) and the proper string partition function.
The first is that in point-particle field theories, we have only thermal
momentum (Matsubara) modes;  there are no analogues
of string thermal winding modes.   
However, while very important in other contexts (\eg, in studies of thermal
duality symmetries~\cite{McClainRoth,AlvOso,AtickWitten,lennek,shyamoli}),
this difference plays no essential role here.
The second difference is the temperature dependence:  Eq.~(\ref{wrongZ}) employs
a standard Boltzman suppression factor, while the thermal components
in Eqs.~(\ref{Zcircdef}) and (\ref{EOfunctions}) are quadratic in
temperature exponential.   Once again, however, this difference is not
relevant in resolving the issue of the missing Hagedorn transition. 
Indeed, upon performing the $\tau$-integration, 
the quadratic temperature dependence 
in Eqs.~(\ref{Zcircdef}) and (\ref{EOfunctions}) 
gives rise to the standard Boltzman suppression factor in the
high-temperature (or high-energy) limits.

The third difference, however, is more significant and goes right
to the heart of string theory.
Using various Schwinger proper-time identities, it is always possible
to recast our field-theoretic partition function into a form
resembling Eq.~(\ref{Vdef});
in such a form,
the quantity $\tau$ emerges as the Schwinger proper time.
However, in this form we would then be instructed to
integrate $\tau$ over
the strip ${\cal S}$ defined as
\beq
    {\cal S}~\equiv~ \left\lbrace
          \tau: ~  
            |\tau_1 |\leq 1/2~,~
        \tau_2 \geq 0 \right\rbrace~.
\label{Sdef}
\eeq
This is to be contrasted with the string calculation, where modular
invariance requires us to restrict our $\tau$-integration 
to the fundamental domain $\calF$ defined in Eq.~(\ref{Fdef}).
This difference has a major effect because the $\calF$ domain
avoids the ultraviolet $\tau_2\to 0$ region completely, 
and this is the region of integration
which ultimately gives rise to the
Hagedorn divergence. 

To see this, note that if we expand each $Z^{(i)}$ in Eq.~(\ref{EOmix}) in
the form of Eq.~(\ref{bare}), we can combine Eqs.~(\ref{spacetimemass})
and (\ref{Zcircdef}) to write $Z_{\rm string}$ in the form
\beq
       Z_{\rm string}(\tau,T) ~=~ {\tau_2}^{(3-D)/2}\,  
       \sum_{\mm,\nn} \sum_{m,n} 
         \,a_{\mm\nn}\, 
      e^{2\pi i\tau_1 (\nn-\mm + mn)}\,
      e^{-\pi \tau_2 \alpha' M_{\mm\nn}^2 }\,
        e^{-\pi \tau_2 \alpha' M_{mn}^2 }
\label{partmass}
\eeq 
where we have separated the ``bare'' zero-temperature mass $M_{\mm\nn}$
in Eq.~(\ref{spacetimemass}) from
the additional ``thermal'' mass $M_{mn}$: 
\beq
         \alpha'\, M^2_{mn} ~\equiv~ {m^2 T^2\over \calM^2 } + {n^2 \calM^2\over T^2}~.
\label{barevstherm}
\eeq
Of course, the $\tau_1$ exponential in Eq.~(\ref{partmass})
is generally trivial:  demanding that $Z_{\rm string}$ be invariant
under the modular transformation $\tau\to\tau+1$ yields the constraint
\beq
           \nn -\mm +mn ~\in~\IZ ~,  
\label{levelmatching}
\eeq
and integrating this exponential 
over $\tau_1$ across the region $-1/2\leq \tau_1\leq 1/2$ enforces
the stronger level-matching constraint $ \nn -\mm +mn =0$ for on-shell string states.
However, our interest is in the behavior of the degeneracies $a_{\nn\nn}$.
Recall from Eqs.~(\ref{asymgrowth}) and (\ref{THag}) that
\beq
           a_{\nn\nn} ~\sim~ e^{M_{\nn\nn}/T_H} ~~~~~ {\rm as}~~\nn\to\infty~, 
\label{asymgrowth2}
\eeq
where we are restricting our attention to $\mm=\nn$ (as would be appropriate,
for example, in the $m=n=0$ thermal sector);
this is the exponential growth that generically leads to a Hagedorn transition. 
However, inserting Eq.~(\ref{asymgrowth2}) into Eq.~(\ref{partmass}) and performing the
sum over $\nn$, we see that the Hagedorn growth in Eq.~(\ref{asymgrowth2}) is 
suppressed by the additional $\tau_2$ exponential in Eq.~(\ref{partmass}) for all $\tau_2>0$.
Indeed, the string partition function diverges only in the $\tau_2\to 0$ region.
Thus, the truncation of the region of integration from $\calS$ to $\calF$ 
in string theory is
ultimately responsible for regulating the ultraviolet behavior and
thereby completely eliminating the Hagedorn transition.
Indeed, this observation has already been made in the 
recent string literature~\cite{shyamoli}.

Note that in this argument, we have been assuming that there are no physical tachyonic states
contributing to Eq.~(\ref{partmass}).
Indeed, the purpose of this argument has merely been to demonstrate that the UV asymptotic
rise in the degeneracy of states does not, in and of itself, lead to a divergence in the
closed-string thermal amplitude. 

It is, of course, always possible as a mathematical exercise to rewrite
our integral over $\calF$ as an integral over $\calS$ using an infinite
set of modular transformations and Poisson resummations~\cite{McClainRoth,Trapletti}.
However, this is merely a mathematical rewriting, and thus cannot introduce
a divergence where none exists.  Indeed, even in such a strip-based
representation for the vacuum amplitudes, there will be delicate algebraic
cancellations which eliminate the supposed UV divergence as $\tau_2\to 0$. 
In the case of the heterotic string which is our focus in this section, 
these cancellations arise through additional phases 
in the strip representation
which emerge as the result of the thermal $\IZ_2$ orbifolding
for worldsheet fermions.

\subsection{The IR argument}

It is also important to understand the elimination of the Hagedorn
transition from the perspective of the infrared argument involving 
low-energy thermal winding
state which becomes massless at the Hagedorn temperature.
However, once again, it is easy to see where the error in this argument
lies.  Recall that this argument makes use of thermal excitations of the
string ground state in Eq.~(\ref{groundstate}).  However, this assumes
that this ground state is actually present   
in the string spectrum --- \ie, that it satisfies the appropriate 
finite-temperature GSO
constraints and appears in the actual finite-temperature string partition function.  
In other words, it is assumed that $Z_{\rm string}$ [or, more precisely,
one of the $Z^{(i)}$ functions in Eq.~(\ref{EOmix})] 
actually contains a term of the form
\beq
         \qbar^{\,-c_R/24} \,q^{-c_L/24}~,
\eeq 
corresponding to the tachyonic string ground state.
Equivalently, phrased in terms of CFT characters of the form
in Eq.~(\ref{Zchi}), it is assumed that $Z_{\rm string}$ contains
a term of the form
\beq
          \overline{\chi}_I\, \chi_I~,
\label{II}
\eeq
representing a tensor product of the identity sectors
of the right- and left-moving worldsheet CFTs.
Unfortunately, this is generally not the case:
in the heterotic string,
this state is always GSO-projected out of the spectrum.
This is certainly true in the zero-temperature theory.
However, our claim is that this is also true in the 
 {\it finite-temperature}\/ theory, even with its modified GSO constraints,
once the correct extrapolation to finite temperature is identified.
Thus, there is no state which survives the GSO projection in order to
trigger the Hagedorn transition.

It is easy to see heuristically why this state must be projected 
out in the case of the heterotic string.
Since the ground
state of the heterotic string has worldsheet energies $(\mm,\nn)=(-1/2,-1)$,
we see from the modular-invariance constraint in Eq.~(\ref{levelmatching})
that such a state could only appear in the term $Z^{(4)}$,
which multiplies the thermal function $\calO_{1/2}$ in Eq.~(\ref{EOmix}).
However, this represents a twisted sector of the $\IZ_2$ thermal orbifold,
and we do not expect to see the ground state of a conformal field
theory emerging from a twisted sector.  
Equivalently stated, we expect a term of the form in Eq.~(\ref{II}) to
appear not within $Z^{(4)}$, but rather from the untwisted sectors
corresponding to $Z^{(1)}$, $Z^{(2)}$, or even $Z^{(3)}$.  
However, modular invariance (specifically invariance under $\tau\to \tau+1$)
prevents this from happening.
Thus, we conclude that this state must be GSO-projected out of the spectrum
in any self-consistent thermal extrapolation of a zero-temperature heterotic
string model.  

This is clearly an important point, and we stress that it requires
a proper understanding of the manner in which a zero-temperature
string model can be self-consistently extrapolated to non-zero temperature. 
While it is certainly consistent {\it at the level of modular invariance}\/
for the CFT ground state $\chibar_I\chi_I$ to appear within
$Z^{(4)}$, our claim is that this is not consistent 
with a proper worldsheet construction for a string theory compactified
on a thermal circle, along with an appropriate $\IZ_2$ orbifold
for fermionic statistics.
Our claim, then, is that any proper finite-temperature extrapolation
will have implicit finite-temperature GSO constraints
which eliminate this state from $Z^{(4)}$.
We shall discuss this conclusion further in Appendix~A, along with explicit
examples. 

Note, in this context, that the two reasons for the absence of
the Hagedorn transition in the heterotic string are actually correlated with each other:
the GSO projections which remove the ground state from the
finite-temperature string spectrum are the same projections which ensure the modular
invariance that allows us to truncate the region of integration
from the strip $\calS$ to the fundamental domain $\calF$ and avoid on-shell tachyons.
Thus, once again, our ultraviolet and infrared Hagedorn arguments
are ultimately correlated.
Essentially, the usual arguments for the Hagedorn transition assume
that the string is nothing more than a tensor product of worldsheet
conformal field theories (CFTs), with a tensor-product ground state
and a tensor-product degeneracy of states.  However, the GSO projections
operate {\it across}\/ the tensor product of CFT's, deforming this structure
into a new collection of surviving states with its own ``effective''
ground state and its own ``effective'' asymptotic degeneracy of states. 

\subsection{The Type~II string}

Finally, let us briefly discuss the case of the Type~II string.
For the Type~II string, the ground state $\overline{\chi}_I\chi_I$ is already level-matched;
thus, the above argument for the heterotic string no longer applies.
In particular, modular invariance and the fact that this state is bosonic would allow 
this term to appear within either $Z^{(1)}$ or $Z^{(3)}$.
However, we are assuming that our Type~II theory is tachyon-free at zero
temperature;  thus this term cannot appear within 
$Z^{(1)}$.
The existence or non-existence of the Hagedorn transition in Type~II strings
thus depends on whether the ground state  
$\overline{\chi}_I\chi_I$ appears within
$Z^{(3)}$.  

To phrase this result somewhat differently,
recall that have already seen below Eq.~(\ref{tildeZ}) that 
any finite-temperature closed string model
can be viewed as interpolating between its zero-temperature version
(presumed supersymmetric) as $T\to 0$ and a 
different, non-supersymmetric zero-temperature string model   
as $T\to \infty$.  This latter model must be non-supersymmetric because we
expect thermal effects to break any supersymmetry which might have existed
at zero temperature.  
Thus, we see that a given Type~II string will have a Hagedorn transition
if and only if it interpolates between a tachyon-free model as $T\to 0$
and a {\it tachyonic}\/ model as $T\to \infty$.    
In other words, the existence of a Hagedorn transition
depends on whether the $T\to\infty$  limit of the Type~II string in question
is tachyonic or tachyon-free.

Unfortunately, this is a model-dependent issue.
In ten dimensions, it is easy to demonstrate that all non-supersymmetric
Type~II string models are tachyonic;  these are the so-called Type~0A and Type~0B
strings.  Thus, 
the only possible $T\to\infty$ endpoints for the Type~IIA/B strings in
ten dimensions are tachyonic, 
leading to the usual Hagedorn transition.
However, in lower dimensions, it is possible to construct Type~II strings
which are non-supersymmetric but tachyon-free.  [Such strings are analogues
of the ten-dimensional non-supersymmetric heterotic $SO(16)\times SO(16)$ string.]
Thus, it may be possible to construct finite-temperature Type~II 
string models in $D<10$ which have such non-supersymmetric,
tachyon-free string models as their $T\to\infty$
endpoints, and for which the Hagedorn transition would be entirely absent.

\section{The Hagedorn transition re-identified}
\setcounter{footnote}{0}

In the previous section,
we demonstrated that the usual Hagedorn transition is 
completely absent in the behavior of thermodynamic quantities
for all heterotic strings, and possibly for certain Type~II strings as well.
Indeed, in such cases,
one-loop thermodynamic
quantities such as the free energy, the internal energy, the entropy, and even
the specific heat remain smooth and undisturbed as a function of temperature,
and cross the Hagedorn temperature without so much as a ripple.

Does this mean that there is no such thing as a Hagedorn transition in such strings?

In this section, we shall demonstrate that
an alternative phase transition often does exist for such strings,
but at a higher temperature.
Moreover, unlike the traditional Hagedorn transition,
this new transition will be completely observable
in the behavior of physical, thermodynamic quantities,
appearing as an actual divergence or discontinuity
in these quantities
as a function of temperature.
Indeed, as we shall demonstrate, this new phase transition
is much weaker than the traditional (first-order) Hagedorn transition,
occurring with an order which depends on the spacetime dimension.

Because the origins of this new phase transition are similar
to those of the traditional Hagedorn transition,
we shall refer to this new phase transition as a ``re-identified''
Hagedorn transition.  However, we believe that in many cases,
this new transition
is indeed the only true Hagedorn transition that such string theories 
experience.

\subsection{Preliminaries}

To see how this new phase transition arises, let us begin more generally by 
determining whether it is {\it ever}\/ possible for our
string thermodynamic quantities to experience divergences or discontinuities.
Although some of the discussion in this subsection is based on 
ideas already presented in previous sections,
our purpose here is to provide a top-down, systematic derivation of the conditions 
for phase transitions in closed-string thermodynamics.

Recall from Sect.~2 that our thermodynamic quantities take the form
\beq
       \int_\calF {d^2 \tau\over \tau_2^2}
    \, \tau_2^x \, \sum_{yz} \, a_{yz} \, \qbar^y q^z~
\label{one}
\eeq
where we have expanded $Z_{\rm string}$ as 
a power series in $q$ and $\qbar$ with temperature-dependent exponents 
$(y,z)$.  Note that 
the modular invariance of $Z_{\rm string}$ 
implies $y-z\in\IZ$.
As a result of the truncation of the region of integration
to $\calF$, there is no ultraviolet divergence from the
region $\tau_2\to 0$.
Thus, the only possible divergence is an infrared
divergence arising from the region $\tau_2\to\infty$.
In this region, however, the $\tau_1$-integration 
across the entire range $-1/2\leq \tau_1\leq 1/2$ enforces level-matching,
eliminating terms in Eq.~(\ref{one}) for which $y\not=z$.
Thus, only terms of the form $\qbar^y q^y$ survive.  
In the infrared region
$\tau_2\to\infty$, our integral therefore behaves as
\beq
       \int^\infty \, {d\tau_2\over \tau_2^2} \,\tau_2^x\, 
           \sum_{y} \,a_{yy}\,   \exp\left( -4\pi \tau_2 y\right)~.
\label{two}
\eeq

Of course, we have already seen in Sect.~3 that 
$a_{yy}\sim e^{\sqrt{y}}$ 
as $y\to\infty$.  
Thus, the $y$-summation in Eq.~(\ref{two}) is convergent  
for all $\tau_2 >0$.    
In other words, we cannot achieve a divergent integral
as a result of the summation over physical string states.

We see, then, that 
divergences in Eq.~(\ref{two}) can arise only if
an individual term in Eq.~(\ref{two}) is divergent.
However, it is immediately clear that all terms with $y>0$ are
individually convergent for all $x$.  Thus, depending on the value of
$x$, divergences
can only arise from terms with $y\leq 0$  ---
 {\it i.e.}\/, terms corresponding to massless or tachyonic
string states.
Of course, this is not a surprise:  the appearance of massless
states at a certain critical temperature is associated with
the emergence of long-range order, which signals the onset of
a phase transition.  Likewise, a tachyonic state signifies
an unstable ground state, thereby triggering the vacuum shift associated
with the phase transition.
However, we now see that there is, indeed, 
no other way in which
our string thermodynamic quantities can diverge.

This is an important point which we shall again emphasize:
the study of a possible Hagedorn transition
reduces to a study of the 
tachyonic and/or massless string states, amounting to an essentially
IR analysis.
In particular, the usual UV arguments are not sufficiently
precise in their standard forms to 
determine whether or not a divergence arises for the vacuum amplitude.
   Of course, the UV and IR approaches are related through Poisson
   resummations and through modular mappings between the fundamental
   domain $\calF$ and the strip $\calS$;  thus the two approaches
   are ultimately equivalent.  Our point, however, is that it is the IR
   analysis which is more direct within the 
   $\calF$-representation for the vacuum amplitudes.

Comparing with Eq.~(\ref{barevstherm}),
we see that the parameter $y$ 
can be identified as $\quarter \alpha' M_{\rm tot}^2$,
where
\beq
   4y ~\equiv~ \alpha' M_{\rm tot}^2 ~=~  
       2 (\mm+\nn) ~+~ {m^2 T^2\over \calM^2} ~+~ {n^2 \calM^2\over T^2}~.
\label{totalmass}
\eeq
Here $(\mm,\nn,m,n)$
respectively represent
the zero-temperature right-moving excitation number,
the zero-temperature left-moving excitation number,
the thermal momentum number,
and the thermal winding number of a given string state.
Recall that for physical states in the $\tau_2\to \infty$ region, 
these quantum numbers are subject to the level-matching constraint
\beq
          \nn - \mm + mn ~=~ 0~.
\label{levelmatch2}
\eeq

Given this, let us classify the different classes of 
massless and/or tachyonic states. 
Since the thermal contributions to 
Eq.~(\ref{totalmass}) are necessarily positive, we see that
there are only four ways of achieving level-matched massless or
tachyonic states:
\begin{itemize}
\item    We can have $\mm=\nn=0$, with $m=n=0$. 
         Such terms correspond to massless states (\eg, the graviton) 
         which appear at zero temperature, and which remain massless at all temperatures.
         We shall refer to such states as ``regular massless'' states.
         All string models contain such states.
\item    We can have $\mm=\nn <0$, with $n=0$, $m$ arbitrary (including $m=0$).
         These terms represent physical tachyons at zero temperature.
         As the temperature is increased,
         the thermal excitations with $m\not=0$ eventually become massless and 
         then massive, while the $m=0$ state remains
         tachyonic for all temperatures. 
         Because of their zero-temperature physical tachyons,
         models containing such terms are uninteresting 
         from a phenomenological perspective,
         and will not be considered further. 
         In some sense, they are already unstable at zero temperature, and
         can be expected to undergo zero-temperature phase transitions 
         which are not thermal in nature.
\item    We can have $\mm=\nn <0$, with $m=0$, $n\not=0$.
         A model containing such a term is not tachyonic at zero temperature.
         However, it contains a tower of massive thermal winding states 
         (corresponding to different values of $n$)  which become massless at 
         specific temperatures and then tachyonic beyond those temperatures.
         Formally, these tachyons then persist for all temperatures beyond
         these critical temperatures, leading to a tachyonic $T\to \infty$ endpoint.  
\item    Finally, we can have $\mm\not= \nn$
         and $\mm+\nn <0$, with both $m\not=0$ and $n\not=0$.  
         As long as Eq.~(\ref{levelmatch2}) is satisfied
         and $\alpha' M_{\rm tot}^2=0$, this will also generally represent 
         a massless physical state at two critical temperatures $T_1$ and $T_2$, with $T_1\leq T_2$.
         [These are the analogues of the temperatures $T^{(\pm)}$ in Eq.~(\ref{Tplusminus}).]
         Such a state is then massive for $T<T_1$ 
         (thanks to the non-zero thermal winding mode),
         and is also massive for $T>T_2$
         (thanks to the non-zero thermal momentum mode).
         Between $T_1$ and $T_2$, the state is tachyonic;   
         however if $\mm=0$ or $\nn=0$, then $T_1=T_2$, thereby eliminating
         the intermediate tachyonic temperature interval.
         Note that in all cases, neither the $T\to 0$ nor the $T\to\infty$ endpoints
         exhibit physical tachyons.

\end{itemize}
We shall refer to states in the final category as
``thermally massless'' at their critical temperatures $T_{1,2}$
because masslessness is achieved at these specific temperatures
as the result of a balance between a 
``bare'' tachyonic mass
and an additional non-zero thermal mass contribution.
It is these ``thermally massless'' states which will be our main focus
in the rest of this section.

Note that much of this discussion mirrors the discussion in Sect.~3,
where we reviewed the traditional arguments for the Hagedorn transition.
As we have seen, the traditional arguments in the Type~II case
rely on the existence of states with 
$(\mm,\nn,m,n)= (-1/2,-1/2,0, \pm 1)$.
Such states are in our third category above,
and only exist for those finite-temperature Type~II strings whose
$T\to\infty$ limits are tachyonic. 
Likewise, the traditional argument in the heterotic case
rests upon the existence of massless states 
with $(\mm,\nn,m,n)=(-1/2,-1,\pm 1/2, \pm 1)$.  Such states 
would have been in our final ``thermally massless'' category
if they had survived the appropriate GSO projections. 
However, as we have already discussed, these states generically
fail to survive the appropriate GSO projections, and thus do not appear
in the one-loop string partition function.

\subsection{Physical, off-shell tachyons:  the ``proto-graviton'' and ``proto-gravitino''}

The question that we face, then, is a simple one:
 {\it in the heterotic string, what is the new, ``effective'' string 
ground state which actually survives the finite-temperature GSO projections?}
In other words, what ``thermally massless'' states actually do exist in 
the thermal string partition function?
Note that in the $T\to 0$ limit in which our thermal $\calE/\calO$ functions
melt away, this ultimately becomes a question about off-shell (but otherwise
physical) tachyons in the zero-temperature theory:  which off-shell tachyons
generically survive the GSO constraints in a given heterotic string model?  
Of course, we are concerned with off-shell tachyons because
we are restricting our attention to states with 
$m+n<0$ but $m\not=n$;  as discussed above, on-shell tachyons would have led 
to instabilities already at zero temperature.

Since we are looking for generic states, let us first consider the only
other generic massless states in the heterotic string, namely those associated with 
the gravity multiplet.   
Recall that in the heterotic string, the graviton is realized 
in the Neveu-Schwarz sector as 
\beq
        \hbox{graviton:}~~~~~~~~~~~~  
             g^{\mu\nu}~\subset ~ \tilde b_{-1/2}^\mu |0\rangle_R ~\otimes~ \alpha_{-1}^\nu |0\rangle_L~
\label{graviton}
\eeq
where $\tilde b_{-1/2}^\mu$ and $\alpha^\nu_{-1}$ are respectively the excitations
of the right-moving worldsheet Neveu-Schwarz
fermion $\tilde \psi^\mu$ and left-moving worldsheet coordinate boson $X^\nu$.
Since the Neveu-Schwarz heterotic string ground state 
has vacuum energies $(H_R,H_L)=(-1/2,-1)$, as in Eq.~(\ref{groundstate}),
the states in Eq.~(\ref{graviton}) are both level-matched and massless, with
$(H_R,H_L)=(\mm,\nn)=(0,0)$.  
These states include the spin-two graviton,
the spin-one antisymmetric tensor field, and the spin-zero dilaton.

In a similar vein, any model exhibiting spacetime supersymmetry  
must also contain the gravitino state, realized in the Ramond sector of 
the heterotic string as
\beq
        \hbox{gravitino:}~~~~~~~~~~~~  
             \tilde g^{\alpha \nu} ~\subset ~ 
         \lbrace \tilde b_{0}\rbrace^\alpha  |0\rangle_R ~\otimes~ \alpha_{-1}^\nu |0\rangle_L~.
\label{gravitino}
\eeq
Here $\lbrace \tilde b_{0}\rbrace^\alpha$ schematically indicates the Ramond zero-mode
combinations which collectively give rise to the spacetime Lorentz spinor index $\alpha$,
as required for the spin-3/2 gravitino state.
   
Regardless of the particular GSO projections, we know that the graviton state
(\ref{graviton}) must always appear in the string spectrum;  likewise, if the
model has spacetime supersymmetry, we know that the gravitino state (\ref{gravitino}) must
exist as well.
However, it is then straightforward to show that this implies that certain additional
off-shell tachyons must also exist in the string spectrum.  Specifically, regardless
of the particular GSO projections, we must
always have a spin-one ``proto-graviton'' state $\phi^\mu$ in the Neveu-Schwarz sector:
\beq
        \hbox{proto-graviton:}~~~~~~~~~~~~  
             \phi^\mu ~\equiv~  \tilde b_{-1/2}^\mu |0\rangle_R ~\otimes~ ~|0\rangle_L~;
\label{protograviton}
\eeq
likewise, if the model is spacetime supersymmetric, we must also have a 
spin-1/2 ``proto-gravitino'' state $\psi^\alpha$ in the Ramond sector:
\beq
        \hbox{proto-gravitino:}~~~~~~~~~~~~  
             \tilde \phi^\alpha ~\equiv~ 
         \lbrace \tilde b_{0}\rbrace^\alpha  |0\rangle_R ~\otimes~ ~ |0\rangle_L~.
\label{protogravitino}
\eeq
Note that these are the same states as the graviton/gravitino, except that in each
case the left-moving bosonic excitation is lacking.  However, it is important to realize
that {\it GSO projections are completely insensitive to 
the presence or absence of excitations of the worldsheet coordinate
bosonic fields}.  Thus, since the graviton is always present,
it then follows that the proto-graviton must also always be present;
likewise, if the model is supersymmetric and the gravitino
is present, then the proto-gravitino must also 
always be present.

While there are many ways to see that the GSO projections must treat the states 
in the gravity multiplet and their proto-counterparts in exactly the same manner,   
it is perhaps easiest to understand this fact from modular invariance.
The heterotic string partition function in $D$ dimensions generally takes the form 
\beq
         Z_{\rm model}(\tau) ~=~ \tau_2^{1-D/2}\,
          ~ { \sum_i \overline{\Theta}_i^{14-D} \Theta_i^{26-D}
         \over \overline{\eta}^{12} \eta^{24} }
\label{Zstringgenformtwo}
\eeq
where the $\Theta$-function numerator schematically
represents lattice sums over compactified momenta and winding
modes,  and where the $\eta$-function denominator represents the contributions from
the bosonic oscillators.  While the GSO projections play a role in determining
which particular combinations of $\Theta$-functions can appear in forming a
modular-invariant numerator,
the $\eta$-function denominators are universally present for all string models and
are modular invariant by themselves.
Indeed, as evident from Eq.~(\ref{Zstringgenformtwo}), they are not even affected
by spacetime compactification.
Therefore, as long as the graviton and gravitino exist in a given supersymmetric
string model, the proto-graviton and proto-gravitino states must also exist
because their contributions are all encoded within the same $\eta$-denominator regardless
of the specific $\Theta$-function numerators.

Thus, we conclude that the proto-graviton and proto-gravitino are two off-shell tachyons
with worldsheet energies $(H_R,H_L)=(0,-1)$ which generically appear in all supersymmetric
heterotic string models.    Moreover, as we shall see,
these are often the effective ground
states of the heterotic string which necessarily survive after the GSO projections 
have been applied.

\subsection{Re-identifying the Hagedorn transition}

Let us now consider the physical effects induced by thermal excitations
of these proto-graviton and proto-gravitino states.
Our first task is to determine the situations in which these states can become ``thermally
massless'' --- \ie, the situations in which they can have massless, 
on-shell thermal excitations.

Our calculation proceeds exactly as in previous sections.
The proto-graviton and proto-gravitino each have worldsheet energies
$(H_R,H_L)=(0,-1)$,
and at non-zero temperature their thermal momentum and winding excitations receive
additional contributions of the form in Eq.~(\ref{thermcontribs}).
Requiring massless, on-shell states with $(H_R^{\rm tot},H_L^{\rm tot})=(0,0)$,
we thus find two solutions at two different temperatures:
\beq
     \cases{  m=1,\phantom{/2} ~n=1, ~a=1  ~&$\Longrightarrow$~    corresponds to $\calO_{0}$~,\cr
              m=1/2, ~n=2, ~a=2  ~&$\Longrightarrow$~   corresponds to $\calE_{1/2}$~.\cr}
\label{twosolns}
\eeq
Note that in each case above, we have also indicated which of the $\calE/\calO$ thermal sums
in Eq.~(\ref{EOfunctions}) 
would give rise to this contribution.   

It is now easy to determine which of these solutions is self-consistent from
a physical standpoint.  Because of its integer momentum number, the first solution must
correspond to a spacetime boson.  Thus, it can only apply for the proto-graviton
rather than the proto-gravitino.  However, 
just like the graviton state from which it is derived,
the proto-graviton should appear within 
the untwisted sector $Z^{(1)}$ in Eq.~(\ref{EOmix}), not the
sector $Z^{(3)}$ corresponding to $\calO_0$.
Thus, the first solution in Eq.~(\ref{twosolns}) cannot be realized for either
of our two proto-states.
The second solution, by contrast, can only apply for the fermionic 
proto-gravitino state because of its half-integer momentum number.
Fortunately, because this solution corresponds to $\calE_{1/2}$ rather than $O_{1/2}$,
it requires the proto-gravitino state to appear exactly where it does appear:
within the untwisted fermionic sector $Z^{(2)}$ rather than the
twisted fermionic sector $Z^{(4)}$.

Thus, we conclude that for 
supersymmetric heterotic string models, the proto-gravitino state has
a thermal excitation with $(m,n)=(1/2,2)$ which becomes ``thermally massless''
at the temperature $a\equiv 2\pi T/M_{\rm string}=2$.
By contrast, the proto-graviton state does not have any potentially massless
thermal excitations.

Because of its non-zero thermal momentum and winding numbers, this thermal
excitation of the proto-gravitino state is extremely massive at both small
and large temperatures, becoming massless only at the specific temperature $a=2$.  
Specifically, we see from Eq.~(\ref{totalmass}) that the mass of this state
is given by
\beq
        \alpha' M_{\rm tot}^2 ~=~ 
                   {a^2\over 4} ~+~ {4\over a^2} ~-~ 2~.
\label{massparabola}
\eeq
Note that unlike the case of the traditional $(m,n)=(\pm 1/2,\pm 1)$ excitation
of the heterotic ground state, the $(1/2,2)$ excitation of the proto-gravitino state
never becomes tachyonic:  this state merely hits masslessness at $a=2$
before becoming massive again at higher temperatures.   
Of course, this result is completely consistent with the fact that the proto-gravitino
state is fermionic, since the existence of a physical fermionic tachyon 
at any temperature would violate Lorentz invariance.
 
However, given that this state never becomes tachyonic,
it is natural to wonder whether this state can ever give rise to a Hagedorn
transition.  Indeed, since no tachyon ever develops, it may appear that our 
thermal vacuum amplitude $\calV(T)$
will never diverge.
It is easy to verify this expectation.
In general, the $(m,n)=(1/2,2)$ thermal excitation of the proto-gravitino state
makes a contribution to $\calV(T)$ given by
\beqn
    \calV(T)  &=& -\half \calM^{D-1}\, 
              \int_\calF {d^2\tau\over \tau_2^2} \, \tau_2^{1-D/2} \sqrt{\tau_2}~
              ~{1\over q} ~ \left\lbrack
                \overline{q}^{(a/2-2/a)^2/4} q^{(a/2+2/a)^2/4} \right\rbrack ~+~...\nonumber\\ 
              &=& -\half \calM^{D-1}\, 
              \int_\calF {d^2\tau\over \tau_2^2} \, \tau_2^{1-D/2} \sqrt{\tau_2}~
                ~e^{2\pi \tau_2}
            ~e^{-\pi \tau_2 (a^2/4 + 4/ a^2)}  ~+~...
\label{Vform}
\eeqn
where we have left the temperature $a\equiv 2\pi T/M_{\rm string}$ arbitrary.
Note that the leading $1/q$ factor in the first line of Eq.~(\ref{Vform})
represents the zero-temperature contribution 
from the proto-gravitino, with $(H_R,H_L)=(0,-1)$,
while the remaining factor in brackets represents the thermal contribution
with $(m,n)=(1/2,2)$.
Likewise, we have carefully recorded all factors of 
$\tau_2\equiv {\rm Im}\,\tau$:  we have
two factors of $\tau_2$ in the denominator from the modular-invariant 
measure of integration;
we have $(1-D/2)$ factors in the numerator from the 
zero-temperature partition function
$Z_{\rm model}$ in Eqs.~(\ref{Zstringgenform}) and 
(\ref{Zstringgenformtwo});
and we have an additional factor $\sqrt{\tau_2}$ in the numerator 
from the Matsubara thermal sums in Eq.~(\ref{Zcircdef}).  
However, at $a=2$, this expression reduces to
\beq
    \calV(T)\biggl|_{a=2}  ~=~ -\half \calM^{D-1}\, 
               \int_\calF {d^2\tau\over \tau_2^2} \, \tau_2^{1-D/2} \sqrt{\tau_2} ~+~...
\eeq
and as $\tau_2\to\infty$, this contribution scales like
\beq
               \int^\infty {d\tau_2\over \tau_2^{(1+D)/2}}~. 
\eeq
This contribution is therefore finite for all $D\geq 2$.
This, of course, agrees with our usual expectation that a massless state does
not lead to a divergent vacuum amplitude in two or more spacetime dimensions.

However, let us now investigate temperature derivatives of $\calV(T)$.
As evident from the second line of Eq.~(\ref{Vform}), 
each temperature derivative $d/dT\sim d/da$
brings down an extra factor of $\tau_2$.  In general, this thereby {\it increases the 
tendency towards divergence of our thermodynamic quantities}\/.

Our results are as follows.
The contribution of this thermally excited proto-gravitino state
to the first derivative $d\calV/da$ is given by
\beq
    {d \calV\over da}  ~=~ \pi \calM^{D-1}\, 
              \int_\calF {d^2\tau\over \tau_2^2} \, \tau_2^{1-D/2} \sqrt{\tau_2}~
               \tau_2 \left(  {a\over 4} - {4\over a^3} \right)\, 
                ~e^{2\pi \tau_2}
            ~e^{-\pi \tau_2 (a^2/4 + 4/ a^2)}  ~+~...,
\label{Vformonederiv}
\eeq
but at the temperature $a=2$ we see that the factor in parenthesis 
within Eq.~(\ref{Vformonederiv}) actually vanishes:
\beq
    {d \calV\over da}\biggl|_{a=2}  ~=~ 0~.
\label{firstderiv}
\eeq
It turns out that this is a general property, 
reflecting nothing more than the fact that the slope of the mass function
in Eq.~(\ref{massparabola}) vanishes at its minimum, as it must.
However, taking subsequent derivatives and evaluating at $a=2$, we find
the general pattern
\beq
            {d^p \calV\over da^p} \biggl|_{a=2} ~=~ 
  \calM^{D-1}\, \int_\calF {d^2\tau\over \tau_2^2} \, \tau_2^{1-D/2} \sqrt{\tau_2} 
           ~ f_p(\tau_2) ~+~...
\eeq
where $f_p(\tau_2)$ for $p\geq 2$ is a rank-$r$ polynomial in $\tau_2$ of the form
\beq
              f_p(\tau_2)~=~ A_p \,\tau_2^r ~+~ B_p \, \tau_2^{r-1} ~+~ C_p \tau_2^{r-2}~...~, 
\eeq
where
\beq
            r ~=~ \cases{ p/2 &  for $p$ even\cr
                          (p-1)/2 & for $p$ odd~,\cr}
\eeq
and where the leading coefficients $A_p$ are positive for 
$p=1,2$~(mod 4) and negative for $p=0,3$~(mod 4), with alternating signs for the lower-order
coefficients $B_p$, $C_p$, {\it etc.}
Given these extra leading powers of $\tau_2$,
we thus find that as a result of the proto-gravitino state,
\beq
            {d^p \calV\over dT^p} ~~~~~\hbox{diverges for}~~~~~ \cases{
             D\leq p&  for $p$ odd \cr
             D\leq p+1 & for $p$ even. \cr}
\eeq
Equivalently, in $D\geq 2$ spacetime dimensions, the proto-gravitino state results
in a divergence that first occurs for $d^p \calV/dT^p$, 
where 
\beq
            p ~=~ \cases{  D & for $D$ even \cr
                           D-1 & for $D$ odd.\cr}
\label{order}
\eeq
This divergence then corresponds to a new, re-identified, 
Hagedorn phase transition.

We stress that it is not merely the masslessness
of this thermally-enhanced proto-gravitino state that
results in this phase transition.  It is  
the fact that this masslessness is achieved {\it thermally}\/,
with non-trivial thermal momentum and winding quanta,
that induces this phase transition.  By contrast, a regular massless state
such as the usual graviton or gravitino does not contribute
to any temperature derivatives of $\calV$.

To interpret the physical ramifications of this new Hagedorn phase transition,
we recall that the free energy $F$
scales with $\calV$ itself, while the internal energy $U$ and entropy $S$ involve
$d\calV/dT$, 
the specific heat $c_V$ involves $d^2\calV/dT^2$, and subsequent temperature
derivatives of $c_V$ involve higher derivatives of $\calV$.
We thus see that $F$ never diverges (it would formally diverge only for $D\leq 1$),
while $U$ and $S$ also never diverge for {\it any}\/ spacetime dimensions
because the first derivative $d\calV/dT$ is always finite, 
as shown in Eq.~(\ref{firstderiv}).
The finiteness of $U$ at the critical temperature suggests that this is
not a limiting temperature, but only the location of a phase transition.
Likewise, $c_V$ and $d c_V/dT$ diverge only for $D\leq 3$, while
$d^2 c_V/dT^2$ diverges for $D\leq 5$, {\it etc.}\/

Thus, for heterotic strings in $D$ dimensions,
we conclude that the proto-gravitino {\it does}\/ give rise to a
Hagedorn transition.  The associated Hagedorn temperature 
is $a=2$ as opposed to the traditional value $a=2-\sqrt{2}$.
Moreover, this re-identified phase transition is generally a very weak 
$p^{\rm th}$-order phase transition, where $p$ is given in Eq.~(\ref{order}). 
However, unlike the traditional Hagedorn transition, this re-identified Hagedorn
transition leaves a {\it bona-fide}\/ imprint in the behavior of string thermodynamic
quantities.
In particular, for $D=4$, this is a fourth-order phase transition in which
$d^2 c_V/dT^2$ diverges, causing 
$d c_V/dT$ to experience a discontinuity, the specific heat
$c_V$ itself to experience a kink, and the internal energy function
to have a discontinuous change in curvature.

\section{Other Hagedorn transitions}
\setcounter{footnote}{0}

In the previous section, we analyzed the Hagedorn transition induced
by the heterotic proto-gravitino state.
As we discussed, this off-shell tachyonic state must always exist 
in a supersymmetric heterotic string model, regardless of the spacetime
dimension.  Thus, the Hagedorn transition we found is completely generic
within the class of supersymmetric heterotic string models. 

Depending on the particular model under study, however, there may be
other off-shell (or even on-shell) tachyons whose thermal excitations
can also give rise to Hagedorn-like transitions.
Clearly, only the transition that occurs with the lowest temperature 
will be the ``true'' Hagedorn transition;  beyond this temperature,
the degrees of freedom of the system can change in a way that eliminates
or modifies all further transitions.
Thus, for completeness, it is also important to study the other 
off-shell (and even on-shell) tachyons
which may exist in such models, and the corresponding
Hagedorn transitions which can potentially arise.

Unlike the situation with the proto-graviton and proto-gravitino,
the tachyonic structure of a given string model is highly model-dependent.
The relevant vacuum energies that may emerge depend crucially on the
particular GSO projections, orbifold twists, and the like.
However, for concreteness, we shall here assume  
only half-integer vacuum energies.  Within a heterotic string
whose ground state vacuum energies are $(H_R,H_L)=(-1/2,-1)$,
this then gives rise to only four classes of possible tachyons, grouped
by their vacuum energies:  $(H_R,H_L)=(-1/2,-1)$, $(0,-1)$, $(0,-1/2)$, and
$(-1/2,-1/2)$.  We shall discuss each of these cases in turn.
As a check on our classification,
we note that there exists a duality symmetry 
\beq
                 a ~\to~ 2/a
\label{symm}
\eeq
under which we merely exchange the roles of $\calO_0$ and $\calE_{1/2}$.
Thus, within each class of tachyon vacuum energies, the solutions for
the corresponding potential Hagedorn temperatures should come in dual pairs
related by Eq.~(\ref{symm}). 

The tachyons in the first class, with  $(H_R,H_L)=(-1/2,-1)$,
are the usual ground-state tachyons of the heterotic string.
As we have discussed in Sect.~3, these off-shell tachyons
would seem to have a massless thermal excitation with $(m,n)=(1/2,1)$
(corresponding to $\calO_{1/2}$),
which would in turn give rise to the traditional Hagedorn transition at
$a= 2-\sqrt{2}$.
(This thermal mode would also be massless at the dual temperature
$a=2+\sqrt{2}$, and tachyonic between these two temperatures).
However, as we have discussed in Sect.~4, 
such states are GSO-projected out of the spectrum,
thereby eliminating the corresponding (traditional) Hagedorn divergence.
 
The tachyons in the second class, with $(H_R,H_L)=(0,-1)$, are
the proto-gravity states we introduced in Sect.~5.
As we discussed, although the proto-graviton state with thermal excitation $(m,n)=(1,1)$ 
would seem to lead to a Hagedorn transition at temperature $a=1$,
the $\IZ_2$ thermal orbifold twist actually eliminates this thermal excitation,
requiring only $m\in 2\IZ$ ({\it i.e.}\/, forcing this state to arise in the
$\calE_0$ sector rather than the $\calO_0$ sector).
Thus, the proto-graviton does not give rise to a Hagedorn transition.
By contrast, as we discussed in Sect.~5,
the proto-gravitino has an allowed massless thermal excitation
with $(m,n)=(1/2, 2)$ giving rise to a Hagedorn transition
at the dual temperature $a=2$.
This, then, is our re-identified Hagedorn transition.
Of course, the proto-gravitino state exists only if our string
model has spacetime supersymmetry at zero temperature.
 
This much has already been discussed in previous sections.
Let us now turn to the tachyons in the third class, with $(H_R,H_L)=(0,-1/2)$. 
These off-shell states have a massless thermal excitation $(m,n)=(1/2,1)$ which
would give rise to a Hagedorn transition at the 
self-dual temperature $a=\sqrt{2}$.
If this is to occur, such states must clearly arise in the twisted
$\calO_{1/2}$ sector
[\ie, within $Z^{(4)}$ in Eq.~(\ref{EOmix})].
Of course, whether these states
exist in such a twisted sector is clearly a model-dependent issue.
However, if they do exist, their $(m,n)=(1/2,1)$ thermal excitations
will be massive for all temperatures other than $a=\sqrt{2}$.
Thus, the corresponding Hagedorn transition in this case will be a weak
one whose order, like that induced by the proto-gravitino, 
depends on the spacetime dimension through Eq.~(\ref{order}).
Note that such tachyons can in principle appear 
not only for heterotic strings, but also for Type~II strings.

Finally, we turn to the $(H_R,H_L)=(-1/2,-1/2)$ tachyons.
Once again, such tachyonic states can arise within both Type~II and
heterotic strings.  For this (on-shell) vacuum energy configuration,
there are four different cases which
must be considered:
\begin{itemize}
\item   First, such tachyons can give rise to massless thermal modes 
           of the form $(m\in \IZ, n=0)$, corresponding to $\calE_0$.
           For $m\not=0$, such thermal excitations are massless at 
           $a=\sqrt{2}/|m|$, massive above this temperature, but tachyonic below it.
           Models containing such states would therefore contain physical, 
           on-shell tachyons
           even at zero temperature [this is the non-thermal $(m,n)=(0,0)$
           tachyon itself], and presumably experience non-thermal
           phase transitions already at zero temperature.  Models
           containing such states are therefore beyond our consideration.
\item   Second, such tachyons can give rise to massless thermal modes 
           of the form $(m=0, n\in 2\IZ)$, again corresponding to $\calE_0$.
           For $n\not=0$, such thermal excitations are massless at the dual temperature 
           $a=|n|/\sqrt{2}$, massive below this temperature, and tachyonic above it.  
           However, because these states must multiply $\calE_0$, they can only
           arise in a model which also contains the $(m=0,n=0)$ non-thermal excitation. 
           This is a non-thermal state which is tachyonic at all temperatures,
           including zero.  Therefore, as in the case above, models
           containing such states are beyond our consideration.
\item  Third, such states could in principle give rise to massless thermal
           modes of the form $(m\in \IZ+1/2,n=0)$, corresponding to
           $\calE_{1/2}$.  As in the first case above, such thermal excitations would 
           be massless at $a=\sqrt{2}/|m|$, massive above this temperature, 
           and tachyonic below it.  However, such states can never arise,
           since their non-integer thermal momentum number indicates that
           they must be spacetime fermions.  Lorentz invariance precludes the
           existence of fermionic tachyons.
            (In string language, this happens because the Ramond zero-mode vacuum
             is never tachyonic.)
\item  Finally, such states can give rise to massless thermal modes
           of the form $(m=0, n\in 2\IZ+1)$, corresponding to $\calO_0$.
           Such thermal excitations are massless at the dual temperature 
           $a=|n|/\sqrt{2}$, massive below this temperature, and tachyonic above it.  
           These models are thus tachyon-free at zero temperature,
           but have physical, on-shell tachyons as $T\to \infty$.
           Such models are not inconsistent.  Indeed, the zero-temperature
           Type~II strings in $D=10$ are in this class, with the $(m,n)=(0,\pm 1)$    
           modes giving rise to the traditional Type~II Hagedorn transition
           at $a=1/\sqrt{2}$.  
           However, even for Type~II strings, tachyons
           in this class are not {\it guaranteed}\/ to exist.  In dimensions $D<10$,
           for example, there exist Type~II models which are non-supersymmetric
           but tachyon-free [analogues of the ten-dimensional
           $SO(16)\times SO(16)$ heterotic string model].
           These models could potentially serve as $T\to\infty$ 
           endpoints of a finite-temperature interpolation away from
           a supersymmetric zero-temperature limit, thereby evading
           the traditional Hagedorn transition for such Type~II models.
\end{itemize}

We see, then, that the particular Hagedorn transitions that a given
string model may experience are closely tied to the on-shell and off-shell
tachyonic vacuum structure of the spectrum. 
This in turn depends not only on the original zero-temperature model
in question, but also on the specific interpolation 
that represents its behavior at finite temperature.

\section{Discussion, speculations, and extensions}
\setcounter{footnote}{0}

In this paper, we have critically re-evaluated the Hagedorn transition
within the context of closed, perturbative Type~II and heterotic strings.
As we discussed, the details of the Hagedorn transition
turn out to depend critically on the precise manner in which  
a given zero-temperature string theory is extrapolated to finite 
temperature.  

For broad classes of closed
     string theories, we found that the traditional  
     Hagedorn transition is completely absent when the correct 
     extrapolation is used.  
This is the case for all heterotic strings, 
and also potentially for certain
Type~II strings in $D<10$.
Indeed, within these classes of string theories,
    the usual Hagedorn phase transition leaves
    no imprint whatsoever in the behavior of one-loop thermodynamic
    quantities such as the string free energy, the internal energy,
    or the entropy.
As we explained, the usual Hagedorn transition is eliminated in such 
cases as a result of the GSO projections that are necessary in
order to produce self-consistent
string models at finite temperature.

However, and potentially more importantly, we found 
that there is an alternative ``re-identified'' Hagedorn transition 
for heterotic strings
which is triggered by the thermal winding excitations of a 
different, ``effective'' tachyonic string ground state.
Unlike the usual CFT ground state which is eliminated
by the GSO projections in the heterotic case, this new ``effective'' string
ground state is physical, and serves as the {\it bona-fide}\/ ground state of 
the theory {\it after}\/ the
GSO projections have been applied and after various towers of string states have
been removed.  These new effective string ground states (corresponding, in some cases,
to the proto-gravitino states discussed in Sect.~5) are not as deeply tachyonic 
as the original worldsheet CFT ground states, and consequently their thermal
excitations give rise to Hagedorn-like phase transitions
at higher temperatures than expected.  Nevertheless, in the absence  of the usual Hagedorn
transitions, we believe that these new ``re-identified'' phase transitions are
the only Hagedorn transitions that such strings experience.
 
Let us therefore summarize what we believe to be the final status
regarding the (non-)existence of the Hagedorn transition for closed strings.

For heterotic strings, we believe that the usual Hagedorn transition
does not exist.  Indeed, its existence would require the worldsheet CFT ground state
to appear within the $Z^{(4)}$ sector of the theory, as indicated
in Eq.~(\ref{EOmix}), yet this state should only appear within the untwisted sector.
Thus, barring any counterexamples to this assertion (and we are not aware of any such
special cases), the usual Hagedorn transition cannot exist.  
We shall illustrate this through explicit examples in Appendix~A.
Nevertheless, in certain cases, a ``re-identified'' Hagedorn phase transition may exist.
For heterotic string models which are supersymmetric at zero temperature, 
we showed that there must always exist a re-identified transition 
at higher temperature whose order depends on the spacetime dimension.
However, there may also be other Hagedorn phase transitions which arise at
even lower temperatures, and which would take priority;  the existence
of these other Hagedorn transitions is thus a model-dependent question.
This will be discussed further in Appendix~A.

For Type~II strings, the situation is even more model-dependent.  
As with the heterotic case, the issue boils down to the construction
of self-consistent modular-invariant finite-temperature interpolating models with 
partition functions of the form in Eq.~(\ref{EOmix}).
Such partition functions necessarily interpolate between  the original
model at $T=0$ and a second model which serves as the $T\to \infty$ limit
of the interpolation. 
Since thermal effects necessarily break supersymmetry, the only
finite-temperature partition functions which can describe the
thermal behavior of the Type~IIA and Type~IIB strings 
are those that interpolate between these strings at $T=0$ and
the Type~0A and/or Type~0B strings at $T=\infty$.  Since the latter
strings have $(H_R,H_L)=(-1/2,-1/2)$ tachyons, and since these tachyons
are indeed the superstring CFT ground states,
the ten-dimensional Type~IIA and Type~IIB strings indeed experience 
usual Hagedorn transitions. 
However, this need not hold in $D<10$.
Specifically, for $D<10$ there exist superstring (Type~II) models
which are non-supersymmetric but tachyon-free;  these are analogues
of the ten-dimensional $SO(16)\times SO(16)$ heterotic string.
If such models can be exploited as the $T\to\infty$ endpoints of finite-temperature
interpolations away from zero-temperature supersymmetric Type~II models,
such models would evade the usual Hagedorn transition.
This again is a model-dependent question.

We stress that in cases where our re-identified Hagedorn transition
is triggered by thermal modes that become tachyonic, 
none of the standard interpretations of this transition
need to be modified.
Whether interpreted as a phase transition to a long string
state or as a breakdown of the string worldsheet, our re-identification
merely implies a new temperature for this phase transition
and a different thermal state as its trigger.
On the other hand, the re-identified Hagedorn
transition associated with the proto-gravitino state is always
a weak one whose order depends on the spacetime dimension.  
Indeed, the appropriate thermal excitation of the proto-gravitino
state never becomes tachyonic;  it simply becomes massless at a
single, critical temperature.
In this case, the physical properties of the corresponding phase transition
are undoubtedly different.

Finally, let us briefly mention the case of Type~I strings.  Of course,
our analysis in this paper has focused on only the closed-string sectors;
thus our results should extend directly to the closed-string sectors of the Type~I strings.
However, the open-string sectors are beyond the analysis we have performed.
In particular, the absence of modular invariance and the emergence of
the ultraviolet $t\to 0$ limit of integration for open-string amplitudes
suggests that a Hagedorn transition is quite likely to occur in such sectors.
On the other hand, it is possible that tadpole anomaly constraints can
conspire to eliminate the Hagedorn divergences from such sectors as well,
in analogy with the manner in which modular invariance can eliminate
the Hagedorn divergences from closed-string sectors.
This issue clearly requires further study~\cite{Dbranes,julie,Dudas}.

Of course, our analysis in this paper is subject to a number of important caveats.
First, we are dealing with one-loop string vacuum amplitudes,
and likewise considering only the tree-level (non-interacting) particle spectrum.
Thus, we are neglecting all sorts of particle interactions.
Gravitational effects, in particular, can be expected to change the 
spectrum quite dramatically, and have recently been argued to eliminate
the Hagedorn transition by deforming the resulting spectrum away from the expected
exponential rise in the degeneracy of states.
However, the purpose of this paper has been to show that even in the non-interacting
theory which has been regarded for nearly two decades as the 
classic Hagedorn system, this transition simply does not arise
in the expected way, if at all.
This therefore casts further doubt on the existence of the Hagedorn transition
after interactions are added.

Our results also raise a number of interesting questions.
For example, it would be interesting to understand the phenomenological
consequences of our observations within the recent brane-world scenarios.
In particular, it seems very strange to contemplate a situation in which
the bulk (closed-string) and brane (open-string) sectors give rise to 
different thermodynamic behaviors at finite temperatures, with phase
transitions occurring in one sector but not the other.
Even more fundamentally, one might also consider the ramifications
of these results for the existence of strong/weak coupling dualities at finite temperatures.
If our results are correct,
it would be extremely important to reconcile the co-existence of Type~I 
strings which exhibit Hagedorn transitions at  a certain critical temperature
with heterotic strings (their supposed strong-coupling duals) which either
fail to exhibit any Hagedorn transitions or exhibit only very high-order phase
transitions at very high temperatures.
Likewise, it would also be interesting to extend our results to non-flat backgrounds 
in order to address important questions such as the thermodynamics of black holes,
the AdS/correspondence, and so forth.
In a similar vein, it would also be interesting to understand the interplay between
these results and recent 
studies of thermal duality~\cite{McClainRoth,AlvOso,AtickWitten,lennek,shyamoli},
especially as far as new phase transitions are concerned. 

Another potentially important line of inquiry might be to interpret these results
within the context of a so-called ``misaligned supersymmetry''~\cite{missusy}.  
Misaligned supersymmetry is a general phenomenon which describes the spectrum
of any non-supersymmetric tachyon-free closed string theory, including a
supersymmetric string theory at finite temperature.
One of the implications
of misaligned supersymmetry 
is that while the asymptotic degeneracy of bosonic or fermionic states grows exponentially
with an exponent corresponding to the traditional Hagedorn temperature,
the actual distribution of bosonic and fermionic states
experiences a rapid fluctuation between bosonic and
fermionic surpluses as one proceeds to higher and higher mass levels.
This surprising fluctuation is the manner in which string amplitudes manage to 
remain finite, even without supersymmetry~\cite{missusy,kutasov}.
However, this rapid fluctuation induces a cancellation such that
a new quantity, a so-called ``sector-averaged'' density of states,
experiences only a much {\it slower}\/ exponential growth~\cite{missusy}.   
Indeed, as shown in Ref.~\cite{missusy}, this reduced rate
of exponential growth is directly correlated with the existence of the proto-graviton
and proto-gravitino states which (as we have seen in Sect.~5) give rise to our
re-identified Hagedorn transition with a re-identified Hagedorn temperature.
Thus the true UV manifestation of both the elimination of the traditional
Hagedorn transition and the emergence of a re-identified one at higher temperatures
is likely to be a misaligned supersymmetry in the asymptotic degeneracy of states. 
This too should be further explored.

We see, then, that the issue of the Hagedorn phenomenon in string theory is a subtle
one which depends quite crucially on the manner in which a zero-temperature
string theory is extrapolated to finite temperature.
As such, the Hagedorn temperature --- and indeed the 
entire existence or non-existence of the Hagedorn transition ---
becomes a highly model-dependent question.
This, perhaps, is the most significant lesson to be taken from these results.

\section*{Acknowledgments}
\setcounter{footnote}{0}

This work is supported in part by the 
National Science Foundation
under Grant~PHY/0301998,
by the Department of Energy under Grant~DE-FG02-04ER-41298,
and by a Research Innovation Award from
Research Corporation. 
We wish to thank 
S.~Abel, J.~Erlich, H.~Firouzjahi, E.~Poppitz, H.~Tye,
and especially E.~Dudas for discussions. 
We would also like to thank S.~Chaudhuri for explaining
the contents of Ref.~\cite{shyamoli}.
One of us (KRD) would like to acknowledge the 
hospitality of the Aspen Center for Physics,
and thank the Laboratory for Elementary Particle Physics 
at Cornell University 
for hospitality during a visit in November 2004
during which this work was completed and presented.

\setcounter{section}{0}   
\Appendix{}
\setcounter{footnote}{0}

In the main body of this paper, we showed that the usual Hagedorn transition
fails to appear for all heterotic strings,  
and we laid out a set of possible re-identified Hagedorn transitions which can
generally emerge to replace it.  In the case of heterotic strings which are supersymmetric 
at zero temperature, we showed that there always exists a weak, high-order
re-identified phase transition triggered by a thermal excitation of the proto-gravitino state.
However, as we noted,
this need not necessarily be the ultimate Hagedorn temperature that such models
experience, for another tachyonic state could give rise to a different Hagedorn
transition at an even lower temperature, thus dominating the thermodynamics.
The question as to whether this might occur is unfortunately model-dependent.

In this Appendix, we shall address this question explicitly by analyzing the 
specific cases of the three tachyon-free heterotic
string models in ten dimensions.  Along the way, we shall also disucss what we believe constitute
self-consistent finite-temperature extrapolations of zero-temperature string models.
In this Appendix, we shall focus on the supersymmetric $SO(32)$ and $E_8\times E_8$
heterotic strings as well as the non-supersymmetric $SO(16)\times SO(16)$ string.
In each case, we shall demonstrate that the possibility of additional Hagedorn transitions
is indeed realized, and that in each case there
is an alternative Hagedorn transition which occurs not at $a=2$, but at $a=1/\sqrt{2}$.
Note that this is also the temperature of the (traditional) Hagedorn transition
that emerges for Type~IIA and Type~IIB strings.
Thus, we shall see that {\it all tachyon-free closed string models   
in ten dimensions have a universal Hagedorn temperature}.
We do not know whether this property persists to lower dimensions, as this continues
to be a model-dependent question which rapidly becomes more complicated after compactification.

Let us begin our analysis by focusing on the $D=10$ supersymmetric $SO(32)$ heterotic string.
At zero temperature, this model can be described by the partition function
\beq
         Z_{SO(32)} ~=~ Z^{(8)}_{\rm boson}~
         (\overline{\chi}_V-\overline{\chi}_S)
         \,(\chi_I^2 + \chi_V^2 + \chi_S^2 + \chi_C^2)~
\label{SO32partfunct}
\eeq
where the contribution from the worldsheet bosons is given as
\beq
         Z^{(n)}_{\rm boson} ~\equiv ~ {\tau_2}^{-n/2}\, (\overline{\eta}\eta)^{-n}~,
\eeq
where the contributions from the right-moving worldsheet fermions are written in terms of
the barred characters $\chibar_\ibar$ of the transverse $SO(8)$ Lorentz
group, and where the contributions from the left-moving (internal) worldsheet fermions  
are written as products of the unbarred characters $\chi_i$ of an $SO(16)$ gauge group.
The subscripts $I$, $V$, $S$, and $C$ generally refer to the identify, vector, spinor, and conjugate
spinor representations of the $SO(2n)$ gauge group;
these representations have conformal dimensions $\lbrace h_I,h_V,h_S,h_C\rbrace=
\lbrace 0,1/2,n/8,n/8\rbrace$, and have corresponding characters 
which can be expressed in terms of Jacobi $\vartheta$-functions as
\beqn
 \chi_I &=&  \half\,(\tthree^n + \tfour^n)/\eta^n ~=~ q^{h_I-c/24} \,(1 + n(2n-1)\,q + ...)\nonumber\\
 \chi_V &=&  \half\,(\tthree^n - \tfour^n)/\eta^n ~=~ q^{h_V-c/24} \,(2n + ...)\nonumber\\
 \chi_S &=&  \half\,(\ttwo^n + {\vartheta_1}^n)/\eta^n ~=~ q^{h_S-c/24} \,(2^{n-1} + ...)\nonumber\\
 \chi_C &=&  \half\,(\ttwo^n - {\vartheta_1}^n)/\eta^n ~=~ q^{h_C-c/24} \,(2^{n-1} + ...)~
\label{chis}
\eeqn
where the central charge is $c=n$ at affine level one.
Even though the spinor and conjugate spinor representations are distinct,
we find that $\chi_S=\chi_C$ (due to the fact that $\vartheta_1=0$).
For the ten-dimensional transverse Lorentz group $SO(8)$,
the distinction between $S$ and $C$ is equivalent to relative
spacetime chirality.
Note that the $SO(8)$ transverse Lorentz group
has a triality symmetry under which  
the vector and spinor representations are
indistinguishable.  Thus $\overline{\chi}_V=\overline{\chi}_S$, resulting 
in a (vanishing) supersymmetric partition function in Eq.~(\ref{SO32partfunct}).

This much is standard.
However, in order to understand how this ten-dimensional model behaves at 
finite temperature, we must construct
an appropriate nine-dimensional string model which is capable of representing 
this ten-dimensional string at finite temperature. 
As we discussed in Sect.~2, such a nine-dimensional model must interpolate between
the supersymmetric $SO(32)$ heterotic string at $T\to 0$ (analogous to $R\to\infty$, where
$R$ is the compactification radius for the thermal dimension), and another 
(presumably non-supersymmetric) ten-dimensional heterotic string model as $T\to\infty$
(or $R\to 0$).
In ten dimensions, the space of non-supersymmetric heterotic string models is 
extremely limited:  a complete classification~\cite{KLTclassification}
shows that there are only seven such non-supersymmetric models.
These are:
\begin{itemize}
\item  a tachyon-free $SO(16)\times SO(16)$ model \cite{AGMV,DH};
\item  a tachyonic $SO(32)$ model \cite{DH,SW};
\item  a tachyonic $SO(8)\times SO(24)$ model \cite{DH};
\item  a tachyonic $U(16)$ model \cite{DH};
\item  a tachyonic $SO(16)\times E_8$ model \cite{DH,SW};
\item  a tachyonic $(E_7)^2 \times SU(2)^2$ model \cite{DH};  and
\item  a tachyonic $E_8$ model \cite{KLTclassification}.
\end{itemize}
(In all but the last case, the gauge symmetries are realized at affine
level one.)
Note that only the first of these seven models 
is devoid of physical tachyons.  The remaining six models all have 
on-shell tachyons at $(H_R,H_L)=(-1/2,-1/2)$. 

Thus, there are only seven nine-dimensional interpolating models which can 
potentially represent the thermodynamics of the supersymmetric 
$SO(32)$ string, depending on which of the above models is chosen as the
$T\to \infty$ limit.
However, it turns out that not all of these nine-dimensional models actually exist;  
as expected from the $\IZ_2$ nature of the thermal orbifold, we can build self-consistent 
nine-dimensional interpolating
models only when the $T\to\infty$ endpoint model is a $\IZ_2$ orbifold 
of the original zero-temperature supersymmetric $SO(32)$ model. 

This provides a significant constraint on the remaining possibilities.
One possibility, for example, is to construct a nine-dimensional model interpolating 
between the supersymmetric $SO(32)$ string and the {\it non}\/-supersymmetric $SO(32)$ string.
Note that the non-supersymmetric $SO(32)$ string has the partition function
\beqn
       Z ~=~ Z^{(8)}_{\rm boson}~\times &\biggl\lbrace &
    \chibar_I  \,(\chi_I\chi_V+\chi_V\chi_I) ~+~ \chibar_V \,(\chi_I^2 + \chi_V^2) \nonumber\\
      && -~\chibar_S\, (\chi_S^2 + \chi_C^2) ~- ~\chibar_C \, ( \chi_S\chi_C+\chi_C\chi_S)
          ~\biggr\rbrace~.
\label{nonsusyso32}
\eeqn
Using standard techniques described in Ref.~\cite{Rohm,IT,GinspargVafa,julie,Dudas},
it is then straightforward to construct the unique, self-consistent 
nine-dimensional string model that interpolates between 
Eqs.~(\ref{SO32partfunct}) and (\ref{nonsusyso32}).
Basically, we compactify the $SO(32)$ string on a (thermal) circle,
and then orbifold by ${\cal T} Q$ where ${\cal T}$ is a shift of
half the circumference around the thermal circle and where $Q$ is the
$\IZ_2$ orbifold that produces the ten-dimensional non-supersymmetric $SO(32)$   
string from the ten-dimensional supersymmetric $SO(32)$ string.
In this case, $Q$ is nothing but $(-1)^F$ where $F$ represents
spacetime fermion number;  it is in this manner that $Q$ breaks spacetime
supersymmetry.
This results in a nine-dimensional 
interpolating model with the partition function~\cite{julie}
\beqn
    Z_A ~=~  Z^{(7)}_{\rm boson} \,\times \,\bigl\lbrace ~
    \phantom{+}&\calE_0 &
     \lbrack \chibar_V \,(\chi_I^2 + \chi_V^2)  ~-~  \chibar_S \,(\chi_S^2 + \chi_C^2) \rbrack\nonumber\\
   +&\calE_{1/2}  &
       \lbrack \chibar_V \,(\chi_S^2+\chi_C^2)  ~-~  \chibar_S \,(\chi_I^2+\chi_V^2) \rbrack\nonumber\\
   +&\calO_0 &
      \lbrack \chibar_I \,(\chi_I\chi_V+\chi_V\chi_I) ~-~ \chibar_C \,(\chi_S\chi_C+\chi_C\chi_S) \rbrack\nonumber\\
   +&\calO_{1/2} &
       \lbrack \chibar_I \,(\chi_S\chi_C+\chi_C\chi_S) ~-~ \chibar_C \,(\chi_I\chi_V+\chi_V\chi_I) \rbrack ~~\bigr\rbrace~
\label{interp1}
\eeqn
Note, in particular, that this reproduces Eq.~(\ref{SO32partfunct}) in the $T\to 0$ limit
as well as Eq.~(\ref{nonsusyso32}) in the $T\to\infty$ limit.
Comparing Eq.~(\ref{interp1}) with Eq.~(\ref{EOmix}), it is easy to read off the particular
components $Z^{(1,2,3,4)}$.

Another possibility might be to build an interpolation between the supersymmetric $SO(32)$ string
and the  tachyon-free $SO(16)\times SO(16)$ string.
The latter string has partition function
\beqn
       Z ~=~ Z^{(8)}_{\rm boson}~\times &\biggl\lbrace&
    \chibar_I  \,(\chi_V\chi_C+\chi_C\chi_V) ~+~ \chibar_V \,(\chi_I^2 + \chi_S^2) \nonumber\\
      && -~\chibar_S\, (\chi_V^2 + \chi_C^2) ~- ~\chibar_C \, ( \chi_I\chi_S+\chi_S\chi_I)
          ~\biggr\rbrace~.
\label{sosixteenpartfunct}
\eeqn
The corresponding nine-dimensional interpolating model then has the partition function~\cite{IT,julie}
\beqn
    Z_B ~=~  Z^{(7)}_{\rm boson} \,\times \,\bigl\lbrace ~
    \phantom{+}&\calE_0 &
          \lbrack \chibar_V \,(\chi_I^2 + \chi_S^2)  ~-~  \chibar_S \,(\chi_V^2+\chi_C^2)\rbrack\nonumber\\
   +&\calE_{1/2}  & \lbrack \chibar_V \,(\chi_V^2 + \chi_C^2)  ~-~  \chibar_S \,(\chi_I^2+\chi_S^2)\rbrack\nonumber\\
   +&\calO_0 &   \lbrack \chibar_I \,(\chi_V\chi_C+\chi_C\chi_V) ~-~ \chibar_C \,(\chi_I \chi_S+\chi_S\chi_I)\rbrack\nonumber\\
   +&\calO_{1/2} &
         \lbrack \chibar_I \,(\chi_I\chi_S+\chi_S\chi_I) ~-~ \chibar_C \,(\chi_V\chi_C+\chi_C\chi_V)\rbrack
       ~~\bigr\rbrace~.
\label{interp2}
\eeqn
Note that in this case the orbifold $Q$ also involves a non-trivial Wilson line which is responsible
for breaking the gauge group.

As a final example, we can also construct a nine-dimensional interpolation between the 
supersymmetric $SO(32)$ string and the non-supersymmetric $SO(8)\times SO(24)$ string.  Letting
 $\overline{\chi}$, $\chi$, and
$\tilde \chi$ respectively represent the
characters of the right-moving $SO(8)$ Lorentz group, the left-moving internal
$SO(8)$
gauge group, and the left-moving internal $SO(24)$ gauge group,
we recall that the partition function of the ten-dimensional $SO(8)\times SO(24)$ string is given by
\beqn
     Z &=&  \chibar_I \, (\chi_I \tilde \chi_S + \chi_S \tilde \chi_I )
       ~+~  \chibar_V \, (\chi_I \tilde \chi_I + \chi_S \tilde \chi_S ) \nonumber\\
     &&  ~-~  \chibar_S \, (\chi_V \tilde \chi_V + \chi_C \tilde \chi_C )
         ~-~  \chibar_C \, (\chi_V \tilde \chi_C + \chi_C \tilde \chi_V )~.
\eeqn
This then leads to a unique nine-dimensional interpolating model
with the modular-invariant partition function~\cite{julie}
\beqn
    Z_C ~=~  Z^{(7)}_{\rm boson} \,\times \,\bigl\lbrace ~
    \phantom{+}&\calE_0 &  
           \lbrack \chibar_V \,(\chi_I\tilde \chi_I + \chi_S\tilde\chi_S)  ~-~
          \chibar_S \,(\chi_V\tilde \chi_V + \chi_C\tilde\chi_C)
          \rbrack\nonumber\\
   +&\calE_{1/2}  & 
          \lbrack \chibar_V \,(\chi_V\tilde \chi_V + \chi_C\tilde\chi_C)  ~-~ 
              \chibar_S \,(\chi_I\tilde \chi_I + \chi_S\tilde\chi_S) \rbrack\nonumber\\
   +&\calO_0 & 
          \lbrack \chibar_I \,(\chi_I\tilde \chi_S + \chi_S\tilde\chi_I)  ~-~ 
               \chibar_C \,(\chi_V\tilde \chi_C + \chi_C\tilde\chi_V) \rbrack\nonumber\\
   +&\calO_{1/2} & 
         \lbrack
          \chibar_I \,(\chi_V\tilde \chi_C + \chi_C\tilde\chi_V)  ~-~
          \chibar_C \,(\chi_I\tilde \chi_S + \chi_S\tilde\chi_I)
          \rbrack ~~\bigr\rbrace~.
\label{interp3}
\eeqn

At this stage, either $Z_A$, $Z_B$, or $Z_C$ (or others) 
could potentially describe the finite-temperature behavior   
of the ten-dimensional $SO(32)$ heterotic string.
   (For example, in Ref.~\cite{shyamoli} a different partition function
   is constructed and analyzed.)
However, it is important to stress that these are not 
merely random modular-invariant functions which happen to 
exhibit mathematical interpolations 
away from the $T=0$ supersymmetric $SO(32)$ endpoint.
Rather, these functions actually emerge as 
the partition functions of {\it bona-fide
nine-dimensional string models with an identifiable thermal radius of compactification}\/.  
In other words, they are the partition functions of nine-dimensional models
which are explicitly constructed by compactifying the original zero-temperature
model on a circle, and then implementing the $\IZ_2$ orbifold ${\cal T}Q$. 
Only this can guarantee their internal self-consistency at the level of an appropriate
worldsheet string construction.
Indeed, the only differences between these possibilities correspond to the internal
gauge-group Wilson lines, represented by the differences in the orbifold factors $Q$.

Given these candidate possibilities $Z_A$, $Z_B$, and $Z_C$, certain 
features are already obvious.  First, we observe that the CFT 
ground state $\chibar_I\chi_I\chi_I$ or $\chibar_I \chi_I\tilde \chi_I$ never appears
in $Z^{(4)}$ ({\it i.e.}\/, multiplying $\calO_{1/2}$).
As discussed 
in Sect.~4.2, this is because the $\calO_{1/2}$ sector is a twisted sector, 
while the CFT ground state is necessarily untwisted.    Indeed, this possibility is ruled
by worldsheet self-consistency constraints stemming from the nature of the $\IZ_2$ thermal orbifold.
We conclude, then, that this CFT ground state is always GSO-projected out of the spectrum, which
in turn implies that
none of these potential finite-temperature models 
would experience the traditional heterotic Hagedorn phase transition at $a=2-\sqrt{2}$.
As discussed in Sect.~4.2, we believe that this is a general feature for all finite-temperature
heterotic string models.

Another important feature is that each of these potential candidate functions contains
the proto-graviton state multiplying $\calE_0$ and a proto-gravitino state
multiplying $\calE_{1/2}$.  
As a gauge singlet, the proto-graviton
state is encoded within the contributions $\chibar_V \chi_I^2$
or $\chibar_V \chi_I\tilde\chi_I$, while the proto-gravitino state
is encoded within the contributions $\chibar_S \chi_I^2$ or $\chibar_S\chi_I\tilde \chi_I$.
Thus, as discussed in Sect.~5, each of these potential finite-temperature extensions
of the $SO(32)$ string would give rise to a re-identified, tenth-order Hagedorn transition 
at $a=2$.  Of course, as discussed in Sect.~6, this phase transition may become irrelevant 
if there are other Hagedorn-like phase transitions which emerge at lower temperatures. 

Finally, we need to determine which of these candidate functions actually represents the 
finite-temperature behavior of the $SO(32)$ heterotic string.
Our selection is guided by following criterion.
Clearly, each of the above partition functions
represents a valid nine-dimensional model with an identifiable {\it geometric}\/ 
radius of compactification.  However, we can interpret this radius as a true {\it thermal}\/
radius ({\it i.e.}\/, as a {\it temperature}) only if we impose the additional
constraint that all massless states that appear multiplied by $\calE_0$
(for which the thermal excitations are periodic around the thermal circle)
be spacetime bosons, while
all massless states that appear multiplied by $\calE_{1/2}$ (for which
the thermal excitations are antiperiodic around the thermal circle) be spacetime fermions.
Note, in particular, that we do {\it not}\/ make any demands on the
$\calO_0$ or $\calO_{1/2}$ sectors, since these sectors necessarily have 
non-zero thermal winding modes.
Such stringy states therefore have no field-theoretic limits, and are beyond our usual
expectations.  Likewise, by restricting our attention to only the {\it massless}\/ states 
which multiply $\calE_0$ and $\calE_{1/2}$, we are again focusing 
on only the light states which can emerge in an appropriate 
low-energy field-theoretic limit.  
We shall discuss the role of the Planck-scale states shortly.

Given this additional constraint, we immediately see that only $Z_A$ in Eq.~(\ref{interp1})
has the required properties.
Indeed, recalling the conformal dimensions listed above Eq.~(\ref{chis}), 
we see that terms of the form $\chibar_{V} \chi_V^2$ and $\chibar_S \chi_V^2$
(or $\chibar_V \chi_V \tilde \chi_V$ and $\chibar_S \chi_V \tilde \chi_V$)
contain contributions from massless spacetime bosons and fermions respectively,
yet these terms appear in $Z_A$ and $Z_C$ multiplied
by $\calE_{1/2}$ and $\calE_0$ respectively.
Such massless ``field-theoretic'' states therefore violate finite-temperature
spin-statistics, causing us to reject these possibilities. 
Thus $Z_A$ is the unique self-consistent interpolating model which can represent 
the supersymmetric $SO(32)$ heterotic string
at finite temperature.
 
Given this result, we might be tempted to conclude that the 
supersymmetric $SO(32)$ heterotic string gives rise to a re-identified
Hagedorn transition at temperature $a=2$ stemming from the proto-gravitino
thermal term within $\chibar_S \chi_I^2 \calE_{1/2}$  in $Z_A$.
However, we notice that the partition function
$Z_A$ also contains $(H_R,H_L)=(-1/2,-1/2)$ tachyons arising from the terms
$\chibar_I (\chi_I\chi_V+\chi_V\chi_I)$ multiplying $\calO_0$.
This is to be expected, since these are nothing but the 32 tachyons of the
non-supersymmetric ten-dimensional $SO(32)$ heterotic string 
that emerges in the $T\to\infty$ limit.
As discussed in Sect.~6, these tachyonic states   
give rise to a Hagedorn transition at $a=1/\sqrt{2}$, 
and this supplants the much weaker tenth-order Hagedorn 
transition that would have been induced at higher temperatures by the proto-gravitino. 
Thus, we conclude that the supersymmetric $SO(32)$ heterotic string   
actually has a Hagedorn temperature $a=1/\sqrt{2}$, just as for the Type~IIA and 
Type~IIB strings. 
In both cases, this behavior is induced by a $(H_R,H_L)=(-1/2,-1/2)$ tachyon
which emerges for $a>1/\sqrt{2}$.

Having explicitly performed this analysis for the supersymmetric $SO(32)$ heterotic string,
we can quickly  consider the cases of the ten-dimensional
$E_8\times E_8$ and $SO(16)\times SO(16)$ heterotic strings.
The analysis is completely similar.
In each case, we find that we can build many nine-dimensional interpolating
models away from these zero-temperature limits, but in each case, we find 
that the self-consistent $T\to \infty$ endpoint must be a $D=10$ tachyonic 
heterotic string model.  Since the tachyons in these models
are always at the same energy $(H_R,H_L)=(-1/2,-1/2)$, and since they
all must arise within the $\calO_0$ sector of the interpolation,
each of these models must also have a Hagedorn transition 
at $a=1/\sqrt{2}$.

We conclude, then, that {\it all tachyon-free closed string models 
in ten dimensions share a universal Hagedorn temperature}\/.  This applies
not only to the Type~IIA and IIB strings, but also to the supersymmetric
$SO(32)$, $E_8\times E_8$, and non-supersymmetric $SO(16)\times SO(16)$ heterotic strings
as well.  Even though the heterotic strings have a different CFT ground state than
their Type~II cousins, and even though their asymptotic densities of states
rise more rapidly, their thermal GSO projections eliminate their corresponding
(traditional) Hagedorn divergences at $a=2-\sqrt{2}\approx 0.59$,
leaving behind one of the re-identified Hagedorn divergences at $a=1/\sqrt{2}\approx 0.71$.
These GSO projections thus restore a certain symmetry and universality between
the Type~II and heterotic strings, at least in ten dimensions, and suggest that
perhaps a similar universality might exist in lower dimensions as well.

Our analysis has also yielded another surprise.  
It was, perhaps, already expected from Ref.~\cite{AtickWitten} that states with
non-trivial thermal winding modes
might behave in a counter-intuitive fashion, violating finite-temperature
spin-statistics relations in the $\calO_0$ and $\calO_{1/2}$ sectors.
What is more surprising, however, is that all of our partition-function 
expressions necessarily have apparent thermal spin-statistics violations even for
the states with {\it zero}\/ windings, \ie, states which appear in the $\calE_0$
and $\calE_{1/2}$ sectors.
Fortunately, all of these violations are safely at the Planck scale;
indeed, this was one of the criteria that we employed
when selecting self-consistent interpolations.
However, we now see quite generally that {\it Planck-scale}\/ spin-statistics
violations of this sort appear to be unavoidable, even for zero-winding states;
they are required, in some sense, by modular invariance.
It would be interesting to understand the thermal implications of these states as far as 
Planck-scale physics is concerned.

Finally, we stress again that it is important to actually construct such nine-dimensional
interpolating models realized through {\it bona-fide}\/ orbifolds and compactifications
of our ten-dimensional models.  Only this can guarantee the internal self-consistency
of the resulting partition function, and the existence of an associated worldsheet
formulation.  Otherwise, if our task were merely to construct modular-invariant expressions
that appear to be mathematical finite-temperature interpolations, other possibilities would immediately arise.
For example, if modular invariance were our only criterion, we could have begun again
with our supersymmetric $SO(32)$ heterotic string and constructed a trivial
``interpolation'' of the form
\beqn
    Z ~=~  Z^{(7)}_{\rm boson} \,\times \,\bigl\lbrace ~
         \phantom{+}&\calE_0 &  \chibar_V ~(\chi_I^2 + \chi_V^2 + \chi_S^2 + \chi_C^2) \nonumber\\ 
         -&\calE_{1/2} &  \chibar_S ~(\chi_I^2 + \chi_V^2 + \chi_S^2 + \chi_C^2)\nonumber\\ 
         -&\calO_{0} &  \chibar_C ~(\chi_I^2 + \chi_V^2 + \chi_S^2 + \chi_C^2)\nonumber\\ 
         +&\calO_{1/2} &  \chibar_I ~(\chi_I^2 + \chi_V^2 + \chi_S^2 + \chi_C^2)~\bigr\rbrace~.
\label{fake}
\eeqn
Indeed, this interpolation would appear to have a traditional Hagedorn divergence
stemming from the $\chibar_I\chi_I^2 \calO_{1/2}$ term, and would also apparently
avoid thermal spin-statistics violations at the Planck scale by cleanly separating the 
zero-temperature bosonic and fermionic contributions within Eq.~(\ref{SO32partfunct})
into separate $\calE_0$ and $\calE_{1/2}$ sectors.
However, it is easy to see that Eq.~(\ref{fake}) cannot correspond to a {\it bona-fide}\/ nine-dimensional
string model.  For example, Eq.~(\ref{fake}) would appear to represent a non-supersymmetric
interpolation between
two {\it supersymmetric}\/ limits, one at $T=0$ and the other at $T\to \infty$,
both of which represent the same $SO(32)$ heterotic string model but with opposite
spacetime chiralities!  Unfortunately, there is no self-consistent $\IZ_2$ orbifold 
$Q$ which can accomplish this feat in ten dimensions.
This example thus illustrates the need to have a proper worldsheet
formulation for our partition function expressions before forming conclusions
about their thermal behaviors.
In other words, string consistency appears to {\it require}\/ 
Planck-scale thermal spin-statistics violations, and in so doing eliminates the usual
Hagedorn transition, replacing it with a re-identified Hagedorn transition
at higher temperature.

\bigskip

\vfill\eject

\bibliographystyle{unsrt}

\end{document}